\documentclass[pra,twocolumn,groupedaddress,aps,floatfix]{revtex4-1}

\def\ket#1{\left|#1\right\rangle}

\def\tr{\mathop{\rm Tr}}
\makeatletter
\def\Hy@safe@activestrue{}
\makeatother

\usepackage{amsmath}
\usepackage{amstext}
\usepackage{amsfonts}
\usepackage{flafter}
\usepackage[below]{placeins}
\usepackage{graphicx}
\usepackage{color}
\usepackage{times}

\graphicspath{{Figs/}}

\begin{document}
%\advance\textheight by -1.5in
%\advance\textheight by -2.3in
%\advance\textheight by -4.4in
\title{Dynamically corrected gates for qubits with always-on Ising
  couplings: Error model and fault-tolerance with the toric code}

\author{Amrit De}

\affiliation{Department of
  Physics \& Astronomy, University of California, Riverside, California 92521}

\author{Leonid P. Pryadko}

\affiliation{Department of
  Physics \& Astronomy, University of California, Riverside, California 92521}

%\date{\today}

\begin{abstract}
  We describe how a universal set of dynamically-corrected quantum
  gates can be implemented using sequences of shaped decoupling pulses
  on any qubit network forming a sparse bipartite graph with always-on
  Ising interactions. These interactions are constantly decoupled
  except when they are needed for two-qubit gates. We analytically
  study the error operators associated with the constructed gates up
  to third order in the Magnus expansion, analyze these errors
  numerically in the unitary time evolution of small qubit clusters,
  and give a bound on high-order errors for qubits on a large square
  lattice.  We prove that with a large enough toric code the present
  gate set can be used to implement fault-tolerant quantum memory.
\end{abstract}

\maketitle

%%====================================================================
\section{Introduction}\label{sec.intro}
%%====================================================================

Preserving quantum coherence is key to realizing a quantum computer.
This can be achieved with the help of quantum error
correction\cite{shor-error-correct,Knill-Laflamme-1997,Bennett-1996}
(QEC), which, in theory, enables any size quantum computation as long
as the errors are below certain
threshold\cite{Shor-FT-1996,Steane-FT-1997,Knill-error-bound,Knill-nature-2005,Aliferis-Gottesman-Preskill-2006,Aharonov-Kitaev-Preskill-2006}.
Unfortunately, the threshold to scalable quantum computation is very
stringent, presently around $1\%$ infidelity per local
gate\cite{Dennis-Kitaev-Landahl-Preskill-2002,Raussendorf-Harrington-2007}.
This value corresponds to encoding logical qubits in individual blocks
of physical qubits using toric\cite{kitaev-anyons}, or related
surface\cite{Dennis-Kitaev-Landahl-Preskill-2002} or color
codes\cite{Landahl-2011}, and is a huge improvement over the original
estimates based on concatenated codes where the threshold value can be
two or more orders of magnitude smaller depending on the
details\cite{Knill-error-bound,Rahn-2002,Steane-2003,Fowler-QEC-2004,Fowler-2005,fowler-thesis-2005,Knill-nature-2005,Knill-2005,Aliferis-Gottesman-Preskill-2006,Aharonov-Kitaev-Preskill-2006}.

In spite of this progress, building a quantum computer with hundreds
or thousands of qubits, with gates concurrently operating at the desired level of
accuracy, is a great physics and engineering challenge.  It is being pursued
by a number of groups, using different physical systems for
implementing qubits.  However, the corresponding control algorithms
need not necessarily be developed from scratch, since the different
physical systems share some key properties.

In particular, qubits with always-on couplings are a natural model for
several potential quantum computer (QC) architectures such as the
original Kane proposal\cite{Kane1998}, nitrogen vacancy centers in
diamond\cite{Milburn2010,vanderSar2012}, superconducting phase
qubits\cite{McDermott2005}, and circuit QED
lattices\cite{Koch2010,Majer2007}. When compared to their counterparts with
tunable couplings, qubits with always-on couplings can be expected to
have better parameter stability and longer coherence times. In
addition there is also much to be benefited from over sixty years
of development in nuclear magnetic resonance (NMR) which has resulted
in an amazing degree of control available to such
systems\cite{vandersypen-2004,Criger-2012}.

Related coherent control techniques based on carefully designed pulse
sequences to selectively decouple parts of the system Hamiltonian
have been further developed in application to quantum
computing\cite{viola-knill-lloyd-1999B,jones-1999,Viola-2002,Viola-Knill-2003,khodjasteh-Lidar-2005,Uhrig-2007,Souza-2011}.
While NMR quantum computing is not easily scalable\cite{jones-2001},
it still holds several records for the number of coherently controlled
qubits\cite{Criger-2012}.  However, some of these records have been
achieved with the help of \emph{strongly-modulated} pulses,
computer-generated single- and multi-qubit gates tailored for a
particular system
Hamiltonian\cite{price-1999,price-1999-jmr,price-havel-cory-2000,Fortunato-Cory-smp-2002}.  While such gates can be used in other QC
architectures\cite{Vartiainen-2004}, they may violate scalability.

On the other hand, NMR-inspired techniques like dynamical decoupling
(DD) can also be used to control large systems with local
interactions, where pulses and sequences intended for a large system
can be designed to a given order in the Magnus
series\cite{slichter-book} on small qubit
clusters\cite{sengupta-pryadko-ref-2005,pryadko-sengupta-kinetics-2006}.
DD is also excellent in producing accurate control for systems where
not all interactions are known as one can decouple interactions with
the given symmetry\cite{stollsteimer-mahler-2001,Tomita-2010}.
Moreover, DD works best against errors coming from low-frequency bath
degrees of freedom which tend to dominate the decoherence rates, and
it does not require additional qubits. In short, DD is an excellent
choice for the first level of coherence protection; it's use could
greatly reduce the required repetition rate of the QEC cycle.

This is well recognized in the research community, and applications of
DD for quantum computation are actively investigated by a number of
groups.  However, most publications on the subject illustrate general
principles using just a single qubit as an example, leaving out the
issues of design and simulation of scalable approaches to multi-qubit
dynamical decoupling. While the techniques for larger systems exist,
they typically require longer decoupling
sequences\cite{stollsteimer-mahler-2001,khodjasteh-Lidar-2005,Khodjasteh-Viola-PRL-2009}.

Recently, we have suggested a universal set of high-fidelity one- and
two-qubit gates for any qubit network that forms a sparse bipartite graph with always-on Ising couplings \cite{De-Pryadko-2013}.
These gates have built-in DD-protection against low-frequency phase
noise and use finite-amplitude \emph{shaped pulses} which could be
experimentally implementable. They can be executed in parallel for
different qubits or pairs of qubits.  These features make the
suggested gate set ideal
for implementing QEC with quantum low-density parity check (LDPC)
codes\cite{Postol-2001,MacKay-Mitchison-McFadden-2004}, in particular,
surface codes and their finite-rate
generalizations\cite{Dennis-Kitaev-Landahl-Preskill-2002,Tillich-Zemor-2009,Kovalev-Pryadko-2012}.

%%% In
%%% Ref.~\onlinecite{De-Pryadko-2013}, we presented the results of full
%%% quantum-mechanical simulations of the error-detecting $[[4,2,2]]$
%%% toric code implemented on an Ising chain of five qubits.

In this work we present the details of the gate design, extend
the construction to enable simultaneous gates on a lattice with
unequal couplings, and analyze the error operators associated with the
constructed gates.  Namely we first analyze the errors analytically up to a
cubic order in the Magnus expansion. We further study these errors
numerically by explicitly integrating the Schr\"odinger equation for
time evolution of clusters of up to six qubits, and
give a bound on high-order errors for qubits on a large square
lattice.  Using this bound, we analytically prove that with large
enough toric code the present gate set can be used to implement a
fault-tolerant quantum memory.

The outline of the paper is as follows.  In Sec.~\ref{sec:intro-dd} we
review dynamical decoupling techniques, extend the analytical
perturbation theory\cite{pryadko-quiroz-2007,pryadko-sengupta-2008} to
cubic order, and illustrate it for several single-qubit soft-pulse
sequences.  In Sec.~\ref{sec:gate-set} we present the details of our
universal gate set construction, generalized here to allow
simultaneous two-qubit gates on a bipartite network with unequal Ising
couplings. We analyze the associated error operators both
analytically, to elucidate the dependence on the pulse shapes, and
numerically, by the full unitary evolution of small networks with up to six
qubits. An important result is that for our gates implemented as
second-order pulse sequences, even with very small infidelities,
errors on graphs with vertex degrees exceeding two predominantly
involve three-qubit and larger clusters.  In Sec.~\ref{sec:scaling},
we give a bound on the amplitude of errors forming large clusters on a
square lattice of qubits for gates designed perturbatively, and show
that such gates can be used with the toric code to implement
fault-tolerant quantum memory.  Lastly, we give our conclusions.

\section{Sequence design}
\subsection{Dynamical decoupling basics}
\label{sec:intro-dd}
Dynamical decoupling originates from the Hahn's spin echo
experiment\cite{Hahn-1950}.  In the simplest version, one applies
ideal infinitely-short ``hard'' pulses which perform single-spin
unitary rotations.  Since the corresponding field is infinite, such
rotations are independent of the system Hamiltonian.  When the
integrated pulse amplitude corresponds to a $\pi$ rotation of the
affected spins, the result is a reversal of some terms in the
Hamiltonian.  For example, for a single spin $1/2$ with the
chemical-shift Hamiltonian
\begin{equation}
  \label{eq:chemical-shift}
  H_{\rm CS}={1\over 2}\Delta\, \sigma^z,
\end{equation}
the unitary for a $\pi$ rotation
around the $x$ axis is $P(\pi,\hat{\bf x})\equiv -i\sigma^x$, while
between the pulses the spin experiences free evolution with the
unitary $U_0(t)=\exp(-i t H_{\rm CS})$.   Throughout this paper we use the
standard notation for the
Pauli matrices $\sigma^\mu$, $\mu=x,y,z$.  A
sequence of two such $\pi_x$ pulses applied at $t=t_1$ and $t=t_2$
respectively corresponds to the evolution
\begin{eqnarray}
  U(t)&=&U_0(t-t_2)P(\pi,\hat{\bf x}) \,U_0(t_2-t_1) \,P(\pi,\hat{\bf
    x})\,U_0(t_1)\quad\;\;\nonumber\\ &=&
  -e^{-i(t-t_2)H_{\rm CS}} e^{-i{(t_2-t_1)}\sigma^x H_{\rm
      CS}\sigma^x} e^{-it_1H_{\rm CS}}\nonumber \\
  &=& - e^{-i{(t-2t_2+2 t_1)} H_{\rm CS}},
\label{eq:refocus}
\end{eqnarray}
where we used the Pauli matrix identity $\sigma^x
\sigma^z\sigma^x=-\sigma^z$.  Up to an overall phase, the effect of
the chemical shift is completely suppressed when the interval between
the pulses is a half of the full evolution time, $t=2(t_2-t_1)$.

In practice, the pulse duration cannot be chosen to be arbitrarily short.
For example, in the case of NMR, the chemical shift
Hamiltonian~(\ref{eq:chemical-shift}) is written in the ``rotating
frame,'' the interaction representation with respect to the
Hamiltonian $\hbar\omega_0\sigma^z/2$, where $\omega_0$ is the carrier
frequency of the RF field of the pulse.  The actual pulse must have a
duration greater than a few cycles at this frequency, $\tau_p\agt
2\pi/\omega_0$.  Much more stringent lower limits on the pulse
duration come about when homonuclear addressing is needed---in this
case selectivity can be achieved when the inverse pulse duration is
small compared to the chemical shift difference.

Similar lower limits on the pulse duration $\tau_p$ also exist in the
solid state setting.  For example, in the case of superconducting
phase qubits, the qubits are formed by the two lowest levels of a
non-parabolic potential well.  While the qubit frequency is around
$10^{10}$\,Hz, the need to avoid the $\ket1\to\ket2$ transition
(typically detuned by some $3\%$ of the qubit frequency $\omega_{01}$)
limits\cite{steffen-2003,Motzoi-Gambetta-Rebentrost-Wilhelm-2009} the
pulse duration by $\tau_p\agt 5$\,ns.

Generally, in order for pulse-based control to be effective, the field
of the pulse must dominate the evolution; for the
Hamiltonian~(\ref{eq:chemical-shift}) this implies the requirement
$\tau_p\Delta\alt 1$.  For any finite-amplitude pulse, e.g., described
by the Hamiltonian $H_C={1\over2}V_x(t)\sigma^x$, in the presence of
the chemical shift $\Delta$, the actual rotation occurs around the net
``magnetic field'' vector $\biglb(V_x(t),0,\Delta\bigrb)$.  With
generic pulse shapes (such as a Gaussian), this produces unitary evolution
operator with errors linear in the pulse duration.

The situation gets more complicated in the presence of an environment.  Most
importantly, dynamical decoupling is not effective against relaxation
due to fast degrees of freedom.  For example, in NMR, the nuclear spins
have a large energy splitting $\hbar\omega_{01}$, the relaxation dynamics is
nearly Markovian and is described by the transverse and longitudinal
relaxation times, $T_1\equiv \gamma_1^{-1}$ and $T_2\equiv
\gamma_2^{-1}$.  While hard $\pi$ pulses commute with the relaxation
superoperator, sequences of soft pulses can modify the structure of
the relaxation and in particular, redistribute relaxation rates between
different channels preserving the combination of $2\gamma_1+\gamma_2$ \cite{pryadko-quiroz-2009}.

Dynamical decoupling is much more effective against decoherence caused
by the low-frequency environmental modes.  The corresponding evolution
is commonly modeled by the general Hamiltonian
\begin{equation}
H=H_C+H_0,\quad H_0\equiv H_B +H_S+H_{SB}, \label{eq:general-Hamiltonian}
\end{equation}
where $H_0$ is the Hamiltonian of the qubits and the environment in
the absence of control.  In this work we assume that qubits with
always-on Ising couplings form a bipartite graph $\mathcal{G}\equiv
(V,E)$ with vertex and edge sets $V$ and $E$ respectively.  Namely, we
write the ``system'' Hamiltonian as:
\begin{equation}
  \label{eq:ising-network}
  H_S={1\over2}\sum_{\langle ij\rangle} J_{ij}\sigma_i^z\sigma_j^z,
\end{equation}
where the two points are neighboring (coupling $J_{ij}\neq 0$) if the
corresponding edge is present in the graph $\mathcal{G}$, $(i,j)\in
E$.  We consider decoherence due to slow dephasing of individual qubits,
with the bath and bath-coupling Hamiltonians, respectively,
\begin{equation}
  \label{eq:ising-bath}
  H_B=\sum_i B_i,\quad H_{SB}={1\over2}\sum_i A_i \sigma_i^z.
\end{equation}
We will assume that each qubit has its own individual bath, meaning
that the bath operators $B_j$ commute with each other, and the
coupling operators $A_i$ commute with all $B_j$, $j\neq i$.

For dynamical decoupling to work, the control Hamiltonian $H_C$ must
be dominant.  To this end, we assume that any large energies have
already been eliminated from the system $H_S$ and system-bath coupling
$H_{SB}$ Hamiltonians by going into the corresponding rotating frame
(interaction representation) and keeping only the slow parts.  While
the norm of the bath Hamiltonian $H_B$ needs not be finite, the
evolution it produces in the Hamiltonian $H_{SB}$ must be in some
sense slow.  We will assume an upper limit on the norms of the bath
coupling operators, $\|A_i\|\alt \omega_c$, and also limit the
$p$-times repeated commutators $[B,\ldots,[B,A_i]\ldots ]$ by
$\omega_c^p\|A_i\|$, where $\omega_c$ is the upper cut-off frequency
of the bath.  For a bath of harmonic oscillators (e.g., phonons),
these assumptions imply a cut-off on the allowed occupation number of
each oscillator.  This can be approximated by ensuring that phonon
modes do not form sharp resonances and by providing sufficient
cooling.

The bath model~(\ref{eq:ising-bath}) can be viewed as an effective
description of qubits operating well above the bath frequency cut-off
to eliminate direct spin flip transitions, with dephasing caused by
phonon scattering.  Similarly, the system
Hamiltonian~(\ref{eq:ising-network}) can be generally obtained as an
effective Hamiltonian for any set of couplings as long as the
transition frequencies of the neighboring qubits differ sufficiently.

We also assume the ability to control the qubits individually,
\begin{equation}
  H_C\equiv \sum_i H_{C}^{(i)},\quad H_{C}^{(i)}={1\over
    2}\sum_{\mu=x,y,z} V_{i\mu}\sigma_i^\mu ,
  \label{eq:control-general}
\end{equation}
where the control signals $V_{i\mu}$ are arbitrary, except for some
implicit limits on their amplitude and spectrum.

\subsection{Average Hamiltonian theory}
\label{sec:intro-average-ham}

Generally, the approach is to treat the control
Hamiltonian~(\ref{eq:control-general}) exactly, and analyze the
evolution due to the system~(\ref{eq:ising-network}) and
bath~(\ref{eq:ising-bath}) Hamiltonians using the average Hamiltonian
theory, an improved version of the time-dependent
perturbation theory.  One introduces the exact unitary
\begin{equation}
U_0(t)\equiv T_t\exp\left(-i\int_0^t dt' H_C(t)\right)
\label{eq:control-unitary}
\end{equation}
associated with the control operator, and the interaction
representation
\begin{equation}
  \label{eq:interaction-hamiltonian}
 \tilde
H_i(t)\equiv
\tilde H_S(t) +\tilde H_{SB}(t)+H_B,
\end{equation}
for the remaining parts of the original Hamiltonian, where, for e.g., the
interaction representation of the system
Hamiltonian~(\ref{eq:ising-network}) is
\begin{equation}
\label{eq:ising-network-rotated}
   \tilde H_S(t)\equiv U_0^\dagger(t) H_S U_0(t).
\end{equation}
Then, the entire evolution operator $U(t)\equiv U_0(t)R(t)$ is
decomposed into a product of the unperturbed operator $U_0(t)$ and the
unitary $R(t)$ for the slow evolution which obeys the integral equation
\begin{equation}
  \label{eq:slow-evolution-integral}
R(t)=\openone-i\int_0^t dt' \tilde H_i(t')R(t').
\end{equation}
The equation is formally solved in terms of the time-ordered exponent
\begin{equation}
  R(t)=T_t \exp\left(-i\int_0^t dt'\,\tilde H_i(t')\right);
  \label{eq:R-time-ordered}
\end{equation}
we will also need the corresponding expansion
\begin{eqnarray}
  \label{eq:R-time-ordered-series}
  R(t)
  &=&\sum_{m=0}^\infty{(-i)^m\over m!} T_t \prod_{j=1}^m\int_0^t dt_j \,\tilde
  H_i(t_j).
\end{eqnarray}
The time-ordered exponent (\ref{eq:R-time-ordered}) can also be
rewritten in terms of an average
Hamiltonian\cite{Waugh-Huber-Haeberlen-1968,%
  waugh-wang-huber-vold-1968},
\begin{equation}
  \label{eq:cumulants-defined}
  R(t)\equiv \exp\left(-i t \bar H(t)\right).
\end{equation}
The leading-order term in the expansion $\bar H(t)=\bar H^{(0)}+\bar
H^{(1)}+\ldots$ in powers of  the
 interaction
Hamiltonian $\tilde H_i(t)$ [see
  Eq.~(\ref{eq:interaction-hamiltonian})] is given by its average,
\begin{equation}
  \label{eq:leading-order-general}
  \bar H^{(0)}={1\over t} \int_0^t dt_0 \,\tilde H_i(t_0),
\end{equation}
while higher-order terms are given by multiple time
integrals\cite{BialynickiBirula-Mielnik-Plebanski-1969} of the sums of
commutators of $\tilde H_i(t)$ evaluated at different time moments
$t_j$.  For the order-$m$ average Hamiltonian, $\bar H^{(m)}$, one has
the sum of commutators of $(m+1)$ terms evaluated at time moments
$0\le t_0\le t_1\le \ldots t_m\le t$.  When the interaction
Hamiltonian is a sum of local terms, as $H_0$ in
Eq.~(\ref{eq:general-Hamiltonian}), the average Hamiltonian $\bar H$
can be written as a sum of terms with support on different connected
clusters.  In particular, with the pairwise qubit couplings following
a connectivity graph ${\cal G}$ as in Eqs.~(\ref{eq:ising-network})
and (\ref{eq:ising-network-rotated}), the clusters correspond to
connected subgraphs of ${\cal G}$.  Explicitly, two bonds belong to
the same cluster  if they are connected either directly (i.e., share a
qubit), or via a continuous chain of connected bonds.

Note that, when dealing with the slow bath, it is common to include
the bath Hamiltonian $H_B=\tilde H_B(t)$ as a part of the interaction
Hamiltonian.  It appears unchanged in the leading-order average
Hamiltonian, $\bar H^{(0)}=H_B+\ldots$, while higher order terms of
the expansion contain only multiple commutators of $H_B$ with other
perturbing terms.
%%% Thus, the expansion is valid even though formally
%%% the bath Hamiltonian may have an unlimited norm (e.g., in the case of
%%% harmonic oscillator bath).  The requirement that the norm of the
%%% commutators be limited is much weaker; it can be satisfied, e.g., by
%%% assuming an upper cut-off frequency $\omega_c$ for the frequencies of
%%% the bath oscillators.

\subsection{Average Hamiltonian of a pulse}
\label{sec:intro-average-ham-pulse}

Dynamical decoupling is perturbative in nature.  An analytical
perturbation theory expansion convenient for analyzing the effect of
pulse shaping on the sequences has been constructed by one of us in
Refs.~\onlinecite{pryadko-quiroz-2007,pryadko-sengupta-2008}.  Here we
extend the expansion to include the terms up to third order for the
%single-qubit
spin-in-dephasing-bath Hamiltonian
\begin{equation}
  \label{eq:qubit-bath}
  H_0=B  +A \sigma^z,
\end{equation}
where $A$ and $B$ are $c$-numbers or operators acting on the bath degrees
of freedom.  The one-dimensional pulse (here we assume a rotation
around the $x$-axis) is given by a single-qubit version of the control
Hamiltonian~(\ref{eq:control-general}) with an arbitrary function
$V_x(t)\equiv V(t)$, $0<t<\tau_p$.  The results of this section can be
trivially generalized to a rotation around an arbitrary direction
$\hat{\mathbf{n}}=\hat{\mathbf{x}}\cos\theta+\hat{\mathbf{y}}\sin\theta$
in the $x$-$y$ plane with the help of the unitary $U_\theta\equiv
\openone\cos(\theta/2)-i\sigma^z \sin(\theta/2)$.

The time-dependent perturbation theory is formulated with respect to
the control evolution alone, with the unitary
\begin{equation}
  \label{eq:pulse-U0}
  U_0(t)\equiv \exp\Bigl(-i \int_0^t dt' H_C(t')\Bigr)=e^{-i\phi(t)\sigma^x/2},
\end{equation}
where the time-dependent phase
\begin{equation}
  \phi(t)\equiv \int_0^t dt'\,V(t'). \label{eq:phase}
\end{equation}
If we denote the net rotation angle $\phi_0\equiv \phi(\tau_p)$, in
the case of a symmetric pulse shape, $V(\tau_p-t)=V(t)$, the rotation
angle has the property $\phi(\tau_p-t)=\phi_0-\phi(t)$.  For such
cases it is convenient to introduce the symmetrized rotation angle,
$\varphi(t)\equiv \phi(t)-\phi_0/2$.  This function is odd under the
pulse-reflection symmetry, $\varphi(\tau_p-t)=-\varphi(t)$.

Using the explicit form (\ref{eq:pulse-U0}) of the evolution matrix
due to the pulse, the interaction representation of the spin-in-a-bath
Hamiltonian~(\ref{eq:qubit-bath}) is just a spin rotation
around the $x$ axis,
\begin{equation}
  \label{eq:interacting}
  \tilde H_0(t)\equiv U_0^\dagger H_0U_0
  =B+A (\sigma^z \cos\phi+\sigma^y\sin\phi).
\end{equation}
The ``slow'' evolution is described by the
unitary $R(t)\equiv U_0^\dagger (t) U(t)$ which obeys the equation
\begin{equation}
  \label{eq:slow-evolution}
  i\dot R(t)=\tilde H_0(t) R(t),\quad R(0)=\openone.
\end{equation}

The net evolution over the duration of the pulse is given in terms of
the corresponding average Hamiltonian $\bar H_0$,
\begin{equation}
  \label{eq:single-qubit}
  U(\tau_p)=U_0(\tau_p) R(\tau_p),\quad R(\tau_p)\equiv e^{-i \tau_p \bar H_0},
\end{equation}
where $\bar H_0=\bar H_0^{(0)}+\bar H_0^{(1)}+\ldots$.  Given that
the interaction Hamiltonian $\tilde H_0(t)$ [see
  Eq.~(\ref{eq:interacting})] at time moment $t_j$ is a sum
of constant
operators multiplied by the functions $c_j\equiv \cos\varphi(t_j)$,
$s_j\equiv \sin\varphi(t_j)$, and a constant $e_j\equiv 1$, the average
Hamiltonian can be computed order-by-order for an arbitrary pulse
shape, in terms of the integrals of products of $c_j$, $s_j$, and
$e_j$.

For a symmetric pulse, the only non-trivial coefficient in the
leading order is
\begin{equation}
  \label{eq:upsilon-defined}
  \upsilon\equiv \langle \cos \varphi\rangle=\int_0^{\tau_p}{dt\over
    \tau_p}\cos \varphi(t),
\end{equation}
which gives  the leading-order average
Hamiltonian\cite{pryadko-sengupta-2008}
\begin{equation}
  \bar H_0^{(0)}= B+\upsilon A\Bigl(\sigma^y \sin{\phi_0\over2}+\sigma^z
  \cos{\phi_0\over2}\Bigr).
  \label{eq:zeroth-order-ham}
\end{equation}
NMR-style first-order self-refocusing
pulses\cite{warren-herm,sengupta-pryadko-ref-2005,pryadko-sengupta-2008}
have $\upsilon=0$.

Similarly, there are only two independent coefficients in the next
order,
\begin{eqnarray}
  \label{eq:beta-defined}
  \beta&\equiv& {1\over2\tau_p^2}\int_0^{\tau_p} dt' \int_0^{t'} dt\,
  \sin\biglb(\phi(t')-\phi(t)\bigrb),\\
    \xi&\equiv& \int_0^{\tau_p}
  {dt\over \tau_p}\Bigl({t\over \tau_p}-{1\over2}\Bigr)
  \sin\varphi(t),
  \label{eq:zeta-defined}
\end{eqnarray}
so that the first-order average Hamiltonian reads
%\begin{widetext}
\begin{equation}
  H_0^{(1)}=  \beta\tau_p \sigma^x
  A^2 + i \xi\tau_p [B,A]\Bigl(\cos {\phi_0\over2}
  \sigma^y   - \sin {\phi_0\over2}
  \sigma^z \Bigr).\label{eq:Have1}
\end{equation}
% \end{widetext}
NMR-style second-order
pulses\cite{sengupta-pryadko-ref-2005,pryadko-sengupta-2008} have
$\upsilon=\beta=0$, which guarantees no error to subleading order
with the chemical shift system Hamiltonian~(\ref{eq:chemical-shift}).
More complicated second-order pulses constructed in
Ref.~\onlinecite{Pasini-2008}, in addition, have $\xi=0$,
which suppresses the entire linear-order average
Hamiltonian~(\ref{eq:Have1}).

Finally, in the third order, out of 27 combinations of
$c_i$, $s_i$, and $e_i$ with $i=1,2,3$, there are only five
independent combinations,
\begin{eqnarray}
  \label{eq:coeffs-third}
  \label{eq:delta1}
  \delta_1&\equiv& \langle c_3 e_2 e_1\rangle-{\upsilon\over 8}\\
  \label{eq:delta2}
  \delta_2&\equiv& \langle s_3 s_2 e_1\rangle,\\
  \label{eq:delta3}
  \delta_3&\equiv& \langle c_3 c_2 e_1\rangle,\\
  \label{eq:delta4}
  \delta_4&\equiv& \langle s_3 s_2 c_1\rangle,\\
  \label{eq:delta5}
  \delta_5&\equiv& \langle s_3 c_2 c_1\rangle,
\end{eqnarray}
where, e.g.,
\begin{equation}
  \label{eq:average3-def}
  \langle s_3 c_2 c_1\rangle\equiv
  \iiint\limits_{0<t_1<t_2<t_3<\tau_p}\!\!\!
  {dt_3dt_2dt_1\over\tau_p^3}\sin\varphi_3
  \cos\varphi_2\cos\varphi_1.
\end{equation}
With the Ising system Hamiltonian~(\ref{eq:qubit-bath}), only the
first four coefficients enter the second-order average Hamiltonian:
\begin{eqnarray}
  {\bar H_0^{(2)}}&=&\tau_p^2
  \Bigl(\frac{\upsilon^2}{6}
  - \delta _2-\delta _3\Bigr) [A,[A,B]]\nonumber \\
  & & + \tau_p^2\Bigl(\sigma^y\sin {\phi_0\over2}
  +\sigma^z\cos {\phi_0\over2} \Bigr)\nonumber\\
  & &\quad \times\left\{
    \Bigl({\upsilon\over24}-\delta_1\Bigr)[B,[B,A]]
    -4 \delta _4  \,A^3 \right\}.
  \label{eq:Have2}
\end{eqnarray}
\subsection{Eulerian-cycle dynamical decoupling}
\label{sec:intro-eulerian}

Instead of, or in addition to designing the pulse shapes, one can
compensate evolution errors associated with arbitrary pulse shapes by
designing sequences of such pulses.  At the level of the leading-order
average Hamiltonian, one universal prescription can be formulated
simply in terms of Eulerian cycles on the Cayley graph associated with
the decoupling group\cite{Viola-Knill-2003}.

For a single qubit, up to a phase, the decoupling group is ${\cal
  G}=\{\openone, \sigma^x, \sigma^y,\sigma^z\}$.  It can be generated
by unitaries $g_1$, $g_2$ corresponding to $\pi$ rotations around a
pair of orthogonal directions, e.g., $x$ and $y$ respectively: ${\cal
  G}=\langle g_x,g_y\rangle$.  The corresponding Cayley graph has a
separate vertex for each group element, and directed edges from each
$s\in{\cal G}$ to $s g$, for every group generator $g$.

In notations of the Sec.~\ref{sec:intro-average-ham-pulse}, the two
rotations can be implemented using some pulse shapes $V_x(t)$,
$V_y(t)$, with the nominal rotation angles $\pi$.  Then, the
corresponding real-world unitaries can be written as $U_x\equiv
-i\sigma^x R_x$, $U_y\equiv -i\sigma^y R_y$, where
\begin{equation}
  R_i=\openone +\delta_{i0}+\sigma^x
  \delta_{ix} +\sigma^y \delta_{iy}+\sigma^z \delta_{iz},
  \label{eq:Ri-operators}
\end{equation}
$i=x,y$, and
the errors $\delta_{i\mu}$, $\mu=0,x,y,z$, are a combined result of
the system-bath Hamiltonian $H_0$ and any inaccuracies of the pulse
duration, amplitude, and phase.  The assumption is that the pulses can
be implemented consistently, so that $\delta_{i\mu}$ are the same for
identical pulses applied at different times.

An Eulerian cycle is a sequence of generators (directed edges) such that every
edge of the Cayley graph is visited.  For a single qubit, the sequence can be
chosen, e.g., as $\{g_x,g_y,g_x,g_y,g_y,g_x,g_y,g_x\}$; the corresponding
unitary is given by the product $U^{\rm Euler}=U_x U_y U_x U_y U_y U_x U_y
U_x$.  The key observation\cite{Viola-Knill-2003} is that $U^{\rm Euler}$ does
not contain terms linear in $\delta_{i\mu}$, $\mu\neq0$; this follows from the
fact that the Cayley tree has edges of each type starting from every group
element.  Thus, the leading-order average Hamiltonian $\bar H_0^{(0)}$ is
independent of the spin variables $\sigma^\mu$.

In the notations of Sec.~\ref{sec:intro-average-ham-pulse}, and in the
absence of any pulse
errors [only errors associated with the system-bath
Hamiltonian~(\ref{eq:qubit-bath}) are preserved], we have
\begin{eqnarray}
  \bar H_0^{(0)}&=&B,\\
  \bar H_0^{(1)}&=&i\tau_p{\kappa\over2}(\sigma^x-\sigma^y) [B,A],\\
  {\bar H_0^{(2)}\over \tau_p^2}&=&i\kappa^2\sigma^z[B,A^2]-
  \left({\kappa^2\over4}+\gamma_2+\gamma_3\right)[[B,A],A]\nonumber
  \\
    & & -{\zeta\over2}\sigma^z
  [B,[B,A]],
\end{eqnarray}
where $\kappa\equiv\upsilon\bigr|_{\phi_0=\pi}$, $\zeta\equiv
\xi\bigr|_{\phi_0=\pi}$, and $\gamma_j\equiv
\delta_j\bigr|_{\phi_0=\pi}$, $j=1,\ldots,5$, are defined as the
coefficients in Eqs.~(\ref{eq:upsilon-defined}),
(\ref{eq:zeta-defined}), (\ref{eq:delta1}), \ldots (\ref{eq:delta5})
for the special case of $\pi$ pulses.

Generally, for an $n$-qubit system, the decoupling group has $2n$
generators and dimension $|{\cal G}|=4^n$; thus an Eulerian path
consists of $n2^{2n+1}$ elements.  Because of this exponential
scaling, the Eulerian cycle construction is not directly useful for
large multi-qubit systems\cite{Viola-Knill-2003}.

A generalization of the Eulerian cycle construction which allows to
generate arbitrary gates has been constructed by Khodjasteh and
Viola\cite{Khodjasteh-Viola-PRL-2009,Khodjasteh-Viola-PRA-2009} (more
complicated sequences which allow for cancellation to an arbitrary
order are also available, see
Ref.~\onlinecite{Khodjasteh-Lidar-Viola-2009}).  The main idea is to
construct a non-trivial ``identity'' operator that shares the
leading-order error operators $\delta_{i\mu}$
[cf.~Eq.~(\ref{eq:Ri-operators})] with those of the gate one is trying
to construct.  For a one-dimensional rotation with the pulse shape
$V(t/2)/2$, $0\le t\le 2\tau_p$ (note the stretching and amplitude
reduction), such an identity operator is a combination of the
unstretched pulse and
antipulse\cite{Khodjasteh-Viola-PRL-2009,Khodjasteh-Viola-PRA-2009},
\begin{equation}
  \label{eq:pulse-antipulse}
  V^{\rm(identity)}(t)=\left\{
    \begin{array}[c]{cc}
       V(t),& 0\le t\le \tau_p;\\
      - V(2\tau_p-t),& \tau_p\le t\le 2 \tau_p.
    \end{array}\right.
\end{equation}
Then, if we denote the unitary of the identity operator as $U_I$, and
the unitary of the stretched pulse as $U_V$, the modified Euler
sequence\cite{Khodjasteh-Viola-PRL-2009,Khodjasteh-Viola-PRA-2009}
corresponds to the unitary (total duration $\tau=16\tau_p$):
\begin{equation}
U_V^{\rm Euler}=U_V U_xU_yU_xU_y U_x U_I U_y U_I U_x U_I U_y. \label{eq:dcg-orig}
\end{equation}
If we introduce the unitary corresponding to the ideal gate
$U_V^{(0)}$, and the sequence-error unitary $R_V$, $U_V^{\rm
  Euler}\equiv U_V^{(0)} R_V^{\rm Euler}$, the sequence
(\ref{eq:dcg-orig}) produces $R_V^{\rm Euler}=\openone-16i\tau_p
B+\mathcal{O}(\tau_p^2)$ for any set of pulse shapes implementing the
unitaries in Eq.~(\ref{eq:dcg-orig}).
Alternatively, the leading-order average Hamiltonian of the gate error
is just the bath Hamiltonian, $\bar H_0^{(0)}=B$, independent of the
degrees of freedom associated with the spin being decoupled.

Explicitly, for the system Hamiltonian (\ref{eq:qubit-bath}),
%%% assuming the NMR-style self-refocusing pulses with
%%% $\kappa=\beta=0$,
when symmetric pulse shapes are used to implement the  DCG
corresponding to an angle-$\phi_0$ rotation around the $y$ axis,
the two subleading terms of the average Hamiltonian read
%(\textbf{insert the expression})
\begin{eqnarray}
  \label{eq:dcg1}
\lefteqn{
  {\bar H_0^{(1)}\over\tau_p}= i{\kappa\over2}\sigma^y
  [A,B]+{\beta\over4}\sigma^y A^2-{i\over4}[A,B]} & & \nonumber\\
  & &\qquad \times \left[({2\kappa-\xi C%\cos {\phi_0\over2}
    -2\upsilon S%\sin{\phi_0\over2}
    })\sigma^x
  +({5\upsilon C%\cos {\phi_0\over2}
    -\xi S%\sin{\phi_0\over2}
    })\sigma^z
\right],\quad \\
  \label{eq:dcg2}
  \lefteqn{{\bar H_0^{(2)}\over \tau_p^2}\Bigr|_{\kappa,\upsilon\to0}=
    %%(\mathbf{fix this})
  i\left({\alpha\over4}\sigma^x-{4\alpha+29\beta\over16}\sigma^y\right)
  [A^2,B]}
\\\nonumber &+&
  {\Big(}{29\xi S- 6\delta_1 C-8\zeta\over16}\sigma^z %\\\nonumber & &
+{29\xi C+6\delta_1 S \over 16}\sigma^x {\Big)}[B,[B,A]] \\\nonumber & &
 -{1\over2}\Bigl( \gamma_2+\gamma_3 + {7\over4} (\delta_2 +
 \delta_3)\Bigr)[A,[A,B]] \\\nonumber & &
+{3\delta_4\over2}(S\sigma^x-C\sigma^z)A^3
\label{eq:one-qubit-DCG2}
\end{eqnarray}
where we introduced $C\equiv \cos\phi_0/2$, $S\equiv \sin\phi_0/2$,
and assumed $\kappa=\upsilon=0$ in the second-order effective
Hamiltonian~(\ref{eq:one-qubit-DCG2}).

It is important to note that even though the noise
Hamiltonian~(\ref{eq:qubit-bath}) can formally be decoupled with a
smaller group (e.g., $\{ \openone,\sigma^x\}$), the corresponding
Eulerian DCG would not be sufficient with generic finite-width pulses.
We confirmed this with an explicit calculation for the
partial-group Eulerian-sequence unitary [cf.\ Eq.~(\ref{eq:dcg-orig})]
\begin{equation}
U_V^{\rm Euler'}=U_V U_x U_x U_x U_I U_x. \label{eq:dcg-partial}
\end{equation}
The corresponding effective Hamiltonian gets a correction already in the
leading order:
\begin{eqnarray}
  \label{eq:dcg-1d-ham0}
  {\bar H_0^{(0)}}  = B-{1\over2}\upsilon \sigma^x A\sin(\phi_0/2).
%%{\bf ??(1/3)}
%%  {\bar H_0^{(1)}\over\tau_p}
\end{eqnarray}
This can be compensated by using self-refocusing pulses with
$\kappa=\upsilon=0$.  Then, in the next order we obtain
\begin{eqnarray}
  {\bar H_0^{(1)}\over\tau_p}\Bigr|_{\kappa,\upsilon\to0}&=&
{1\over2}\left({\alpha}\sigma^x+{\beta}\sigma^y\right) A^2\nonumber\\& &
+i {\xi\over2}\left( C\sigma^x+S \sigma^z \right) [A,B].
%\left({2\alpha\over 3}\sigma^x+{\beta\over 6}\sigma^y\right) A^2
%\nonumber\\
%& & +i {\xi\over 6}\left( C\sigma^x+S \sigma^z \right) [A,B].
\label{eq:dcg-1d-ham1}
\end{eqnarray}
This, in turn, can be compensated using the second-order
self-refocusing pulses in which case we are left only with the
second-order Hamiltonian
\begin{eqnarray}
  \nonumber
    \lefteqn{{\bar
        H_0^{(2)}\over\tau_p^2}\Bigr|_{\kappa,\upsilon,\alpha,\beta\to0}
      =  (5\sigma^x S\delta_4 - 3\sigma^z C\delta_4)A^3}& &
%%   i\left({\alpha\over4}\sigma^x-{13\beta\over8}\sigma^y\right)[A^2,B]
  \\\nonumber
 &\quad &
  -\Bigl({\gamma_2+\gamma_3\over2}+5{\delta_2+\delta_3\over4}\Bigr)
   \big[A,[A,B]\big] \\ \nonumber
 &\quad & -\Biggl(\Bigl ({11\xi C\over8}+{5\delta_1 S\over4}\Bigr)\sigma^x
  - \Bigl({\zeta\over2}-{13\xi S\over8} +
  {3\delta_1C\over4}\Bigr)\sigma^z\Biggr)\quad \\
 & & \qquad  \times [B,[B,A]]
  \label{eq:dcg-1d-ham2}
\end{eqnarray}
Note that the pulse shapes from Ref.~\cite{Pasini-2008} have
$\upsilon=\beta=\zeta=0$ ($\kappa=\alpha=\xi=0$ for $\phi_0=\pi/2$);
use of such pulses completely suppresses the subleading
Hamiltonian~(\ref{eq:dcg-1d-ham1}).  In Eqs.~(\ref{eq:dcg-1d-ham1}),
(\ref{eq:dcg-1d-ham2}) we kept $\xi$ and $\zeta$ non-zero, as for
NMR-style self-refocusing
pulses\cite{warren-herm,sengupta-pryadko-ref-2005,pryadko-sengupta-2008}.

\section{Universal gate set for bipartite Ising lattices}
\label{sec:gate-set}

In this section we continue using the Hamiltonian specified by
Eqs.~(\ref{eq:general-Hamiltonian}), (\ref{eq:ising-network}),
(\ref{eq:ising-bath}), and (\ref{eq:control-general}).  An important
property of this Hamiltonian is that even in the presence of control on
non-neighboring qubits (e.g., one of the sublattices) it separates
into small commuting pieces.  These include a ``tuft'' Hamiltonian for
every controlled qubit $j$: a combination of on-site bath coupling
Hamiltonian~(\ref{eq:ising-bath}) with index $j$ and all of the nodes
(\ref{eq:ising-network}) from that vertex.  It is easy to see that the
corresponding single-tuft unitary can be expressed in terms of the
single-qubit average Hamiltonian, see Eqs.~(\ref{eq:zeroth-order-ham}),
(\ref{eq:Have1}), and (\ref{eq:Have2}) for the first three orders.
\subsection{Single-qubit operations}
\label{sec:dcg}
We construct the single-qubit rotations using a version of the
partial-group Eulerian path construction, see
Sec.~\ref{sec:intro-eulerian}.  The qubits are separated into four
groups: idle qubits on sublattices $A$ and $B$, and the qubits on the
same two sublattices which we want to rotate.  These latter should not
neighbor each other.  In a typical application, one-dimensional
rotations can be applied to every qubit of one of the sublattices, $A$
or $B$.

\begin{figure}[htbp]
  \centering
  \includegraphics[width=0.8\columnwidth]{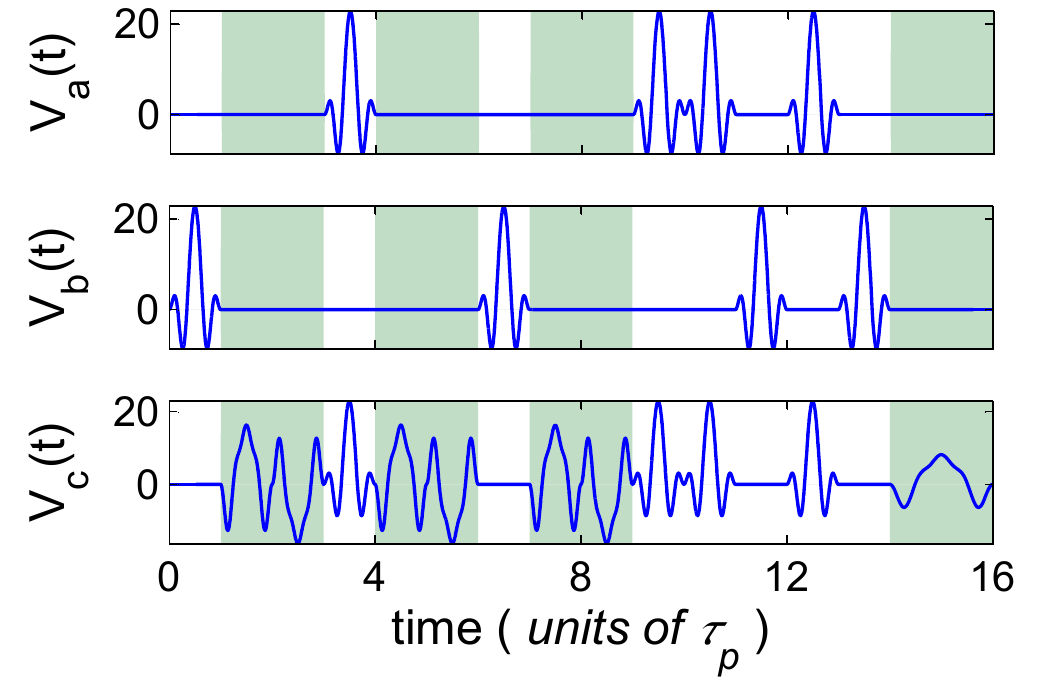}
  \caption{(Color online) An example for executing a single qubit $\pi/2$
    rotation along the $y$-axis(shaded region) using a DCG construction on a bipartite
    lattice such as a star-graph. One or both of the sequences of $\pi$
    pulses along $x$, $V_a(t)$ and $V_b(t)$, are executed globally on the
    idle qubits of the two sublattices.  The single-qubit $(\pi/2)_Y$
    rotation is implemented as a DCG by adding three pulse-antipulse combinations and the stretched pulse along the $y$-axis to the sequence $V_a(t)$ or $V_b(t)$ depending on the sublattice ($V_c(t)$ is executed on sublattice ``a'').  The pulses in the shaded regions are
    $Q_1(\pi/2)$ and the pulses along $x$ are $Q_1(\pi)$ from Ref.~\onlinecite{pryadko-sengupta-2008}.}
  \label{fig:Y3p}
\end{figure}

The sequence is illustrated in Fig.~\ref{fig:Y3p}.  The entire
sequence lasts $\tau=16\tau_p$, where $\tau_p$ is the nominal
single-pulse duration, with the entire interval split into sixteen
equal intervals of duration $\tau_p$.  For the idle qubits on
sublattice $A$, four identical symmetric $\pi_x$ pulses are executed
during the intervals $4, 10, 11$, and $13$ [the top plot, $V_a(t)$, in
  Fig.~\ref{fig:Y3p}].  For the idle qubits on sublattice $B$, the
$\pi_x$ pulses of the same shape are executed during the intervals
$1,7,12$, and $14$, see $V_b(t)$ in Fig.~\ref{fig:Y3p}.  On the
controlled qubits, additional pulses are inserted during the remaining
intervals: a symmetric pulse $V(t)$ during the intervals $2,5,8$, the
same but inverted pulse $-V(t)$ during the intervals $3$, $6$, $9$,
and double-duration half-amplitude pulse $V(t/2)/2$ during the
intervals $15,16$.  All of these pulses should be applied in the
direction of the desired rotation.  The curve $V_c(t)$ in
Fig.~\ref{fig:Y3p} illustrates a $(\pi/2)_Y$ rotation on a qubit of
the sublattice $A$.

The average Hamiltonian corresponding to such a sequence depends on
the chosen graph ${\cal G}$ and on the direction of the applied pulse.
For an open four-qubit chain, the desired rotation around the $Y$ axis
for qubits 1 and 3, and assuming all pulses are symmetric, the
leading-order average Hamiltonian reads
\begin{equation}
  \label{eq:DCG-Y3p-ham0}
  {\bar H_0^{(0)}}  = B-{1\over2}\upsilon
  \sin(\phi_0/2)\sum_{i=1,3}\sigma_i^x A_i.
\end{equation}
This is similar to the case of the partial-group single-qubit DCG, see
Eq.~(\ref{eq:dcg-1d-ham0}): to achieve leading-order decoupling, one
needs to use NMR-style self-refocusing pulses with $\upsilon=0$ like
those developed in
Refs.~\onlinecite{warren-herm,sengupta-pryadko-ref-2005,%
  pryadko-sengupta-2008}.  The first-order average Hamiltonian is a
lengthy expression containing the coefficients $\upsilon$, $\beta$,
$\xi$ (corresponding to the angle-$\phi_0$ pulses) and two of their
counterparts for the $\pi$ pulses, $\kappa$ and $\alpha$.
Unfortunately, the first-order average Hamiltonian remains non-zero
even when the second-order pulses similar to those constructed in
Ref.~\onlinecite{Pasini-2008} are used, with $\upsilon=\beta=\xi=0$,
as well as the regular NMR-style second-order $\pi$ pulses with
$\kappa=\alpha=0$.  When such pulses are used, we have
\begin{eqnarray}
  \label{eq:DCG-Y3p-ham1}
  {\bar H_0^{(1)}
}\Bigr|_{\kappa=\alpha=\upsilon=\beta=\xi=0}
=i{\tau_p\over4}\sum_{i=1}^4\sigma_i^z[B_i,A_i],
\end{eqnarray}
where we used the assumption $[A_i,B_j]=0$, $[A_i,A_j]=0$ for $i\neq
j$.

In order to suppress such error terms, one can use a symmetrized
version of the sequence.  Namely, the pulses in Fig.~\ref{fig:Y3p}
are first executed in reverse order, then directly, for the total
duration of $32\tau_p$.  Since the desired rotation is repeated two
times, the two $\phi_0$ pulses in the symmetrized DCG sequence produce a
rotation of $2\phi_0$.  The corresponding leading-order average
Hamiltonian is just $ {\bar H_0^{(0)}}=B$, while in the first
order (when using the second-order pulses with
$\kappa=\upsilon=\alpha=\beta=0$), the average Hamiltonian is
proportional to $\xi$,
\begin{eqnarray}
  \nonumber
  {\bar H_0^{(1)}
  }\Bigr|_{\kappa=\alpha=\upsilon=\beta=0}&=&i{\tau_p\xi\over4}C
  \Bigl\{(C_2\sigma_1^x+S_2 \sigma_1^z)[A_1,B_1]\\
& & \quad  +(C_2\sigma_3^x+S_2\sigma_3^z)[A_3,B_3]\Bigr\},%%% \\
  %%% & & +i{\tau_p\xi\over4} CS_2
  %%% \left(\sigma_1^z[A_1,B_1]+\sigma_3^z[A_3,B_3]\right),
  \label{eq:DCGi-Y3p-ham1}
\end{eqnarray}
where $C\equiv \cos(\phi_0/2)$, $S\equiv \sin(\phi_0/2)$ as before,
and $C_2\equiv \cos\phi_0$, $S_2\equiv \sin \phi_0$.

\subsection{{$\boldsymbol{ZZ}$} rotation}
\label{sec:zz}
With Ising couplings, the natural two-qubit gate is the $ZZ$ rotation,
$\exp({-i\alpha \sigma^z\otimes\sigma^z})$.  To implement such a gate
between two neighboring qubits on a bipartite lattice with always-on
Ising couplings, one just has to suppress the unwanted couplings.  We
design the corresponding sequences starting first with the sequences
of hard pulses.

\begin{figure*}[htbp]
  \centering
    \includegraphics[width=1.6\columnwidth]{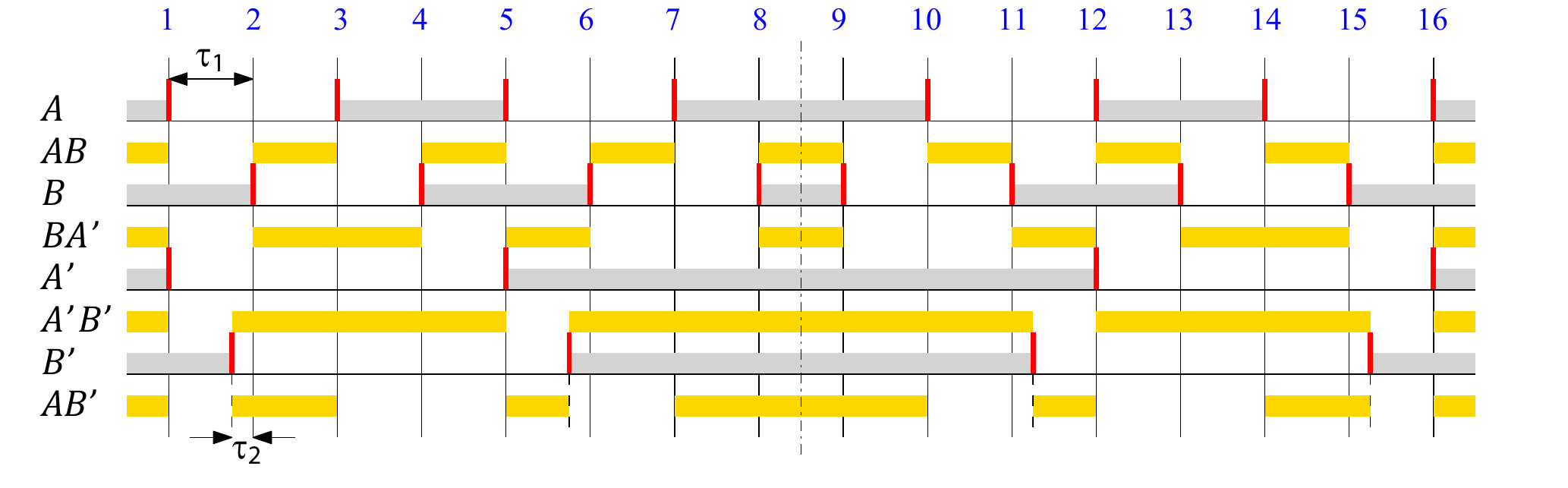}
    \caption{(Color online) Schematic design of the $ZZ$-rotation gate
      on a bipartite Ising network using $\delta$-pulses.  Pulses are
      indicated with vertical red lines (all of them are $\pi$ pulses
      around the $x$-axis).  Sequences $A$ and $B$ are applied on idle
      qubits of the two sublattices.  The regions shaded in gray
      correspond to time intervals where the signs of $\sigma^z$ on
      the corresponding sublattice is not inverted, while yellow shading
      along the intermediate line labeled $AB$ represents the sign of
      the coupling term $\sigma^z\otimes\sigma^z$. All of these
      occupy exactly half of the total cycle duration, indicating that
      the corresponding leading-order average Hamiltonians are all
      zero.  The lines labeled $A'$ and $B'$ correspond to a pair of
      qubits to be coupled.  They are decoupled both from the on-site
      noise and from the neighboring dual-sublattice qubits as can be
      seen from the shading along lines labeled $A'$, $B'$, $AB'$, and
      $BA'$.  On the other hand, the mutual coupling (line $A'B'$)
      does not average to zero, see Eq.~(\ref{eq:avhamAB}).}
  \label{fig:seq}
\end{figure*}

Consider two doubled partial-group Eulerian sequences, each
constructed as four equally spaced $\pi_x$ pulses, followed by an
exactly reversed sequence, see lines $A$ and $B$ in
Fig.~\ref{fig:seq}.  Taking the time interval between the pulses to be
$\tau_1$ (see Fig.~\ref{fig:seq}), the $A$ sequence has first four
pulses centered at the odd-numbered intervals of duration $\tau_1$
(intervals 1, 3, 5, 7), and the trailing four pulses centered at
even-numbered intervals (10, 12, 14, 16), for the total sequence
duration $\tau=16\tau_1$.  The $B$ sequence has this pattern reversed,
with pulses centered at intervals 2, 4, 6, 8, 9, 11, 13, 15.  These
sequences provide decoupling of both the single-qubit and the Ising
Hamiltonians, see Eqs.~(\ref{eq:ising-bath}) and
(\ref{eq:ising-network}), as can be deduced from the shading in lines
$A$, $B$, and $AB$ in Fig.~\ref{fig:seq}.  Due to the sequence
symmetry, with $\delta$-pulses, all odd orders in the Magnus series
are suppressed, which guarantees the second order cancellation.

Now, any similarly-constructed double-interval sequence (e.g.,
sequence $B'$ in Fig.~\ref{fig:seq}) decouples the corresponding qubit
from those on lines $A$ and $B$, and also provides the decoupling of
the single-qubit Hamiltonian~(\ref{eq:ising-bath}).  We use this
freedom to construct sequences $A'$ and $B'$ that provide
continuously-varied coupling:
\begin{equation}
  \label{eq:avhamAB}
  {\bar H^{(0)}}_{A'B'}=f\,{1\over2}J_{12}\sigma^z_1\sigma^z_2,\quad
  f={8\tau_1+8\tau_2\over
    16\tau_1},
\end{equation}
where the prefactor $f$ is the result of the averaging.  With ideal
$\delta$ pulses and no dead-time intervals, the time shift between the
sequences must satisfy the condition $-\tau_1\le \tau_2\le \tau_1$;
this gives full control over values of the prefactor, $0\le f\le 1$.
Note also that the leading-order average
Hamiltonian~(\ref{eq:avhamAB}) becomes exact to all orders, $ {\bar
  H}_{A'B'}= {\bar H^{(0)}}_{A'B'}$, when the bath couplings [see
  Eq.~(\ref{eq:ising-bath})] are replaced with time-independent energy
shifts, $A_i\to \Delta_i$, or when the individual bath Hamiltonians
are dropped, $B_i\to0$.

When $\delta$-pulses are replaced with soft pulses of duration
$\tau_p$ centered at the same positions, the corresponding
leading-order average Hamiltonian remains parametrically the same, see
Eq.~(\ref{eq:avhamAB}).  However, since the allowed range of the
time shift must be reduced to avoid pulse overlaps, $|\tau_2|\le
\tau_1-\tau_p$, the prefactor  $f$ can only be tuned
in the range $\tau_p/2\tau_1\le f\le 1- \tau_p/2\tau_1$.  When used
with the NMR-style second-order pulses ($\kappa=\alpha=0$), the
first-order average Hamiltonian is zero, while the second-order
average Hamiltonian is a complicated expression depending on the
graph that describes the inter-qubit couplings.

\begin{figure}[htbp]
  \center
  \includegraphics[width=0.9\columnwidth]{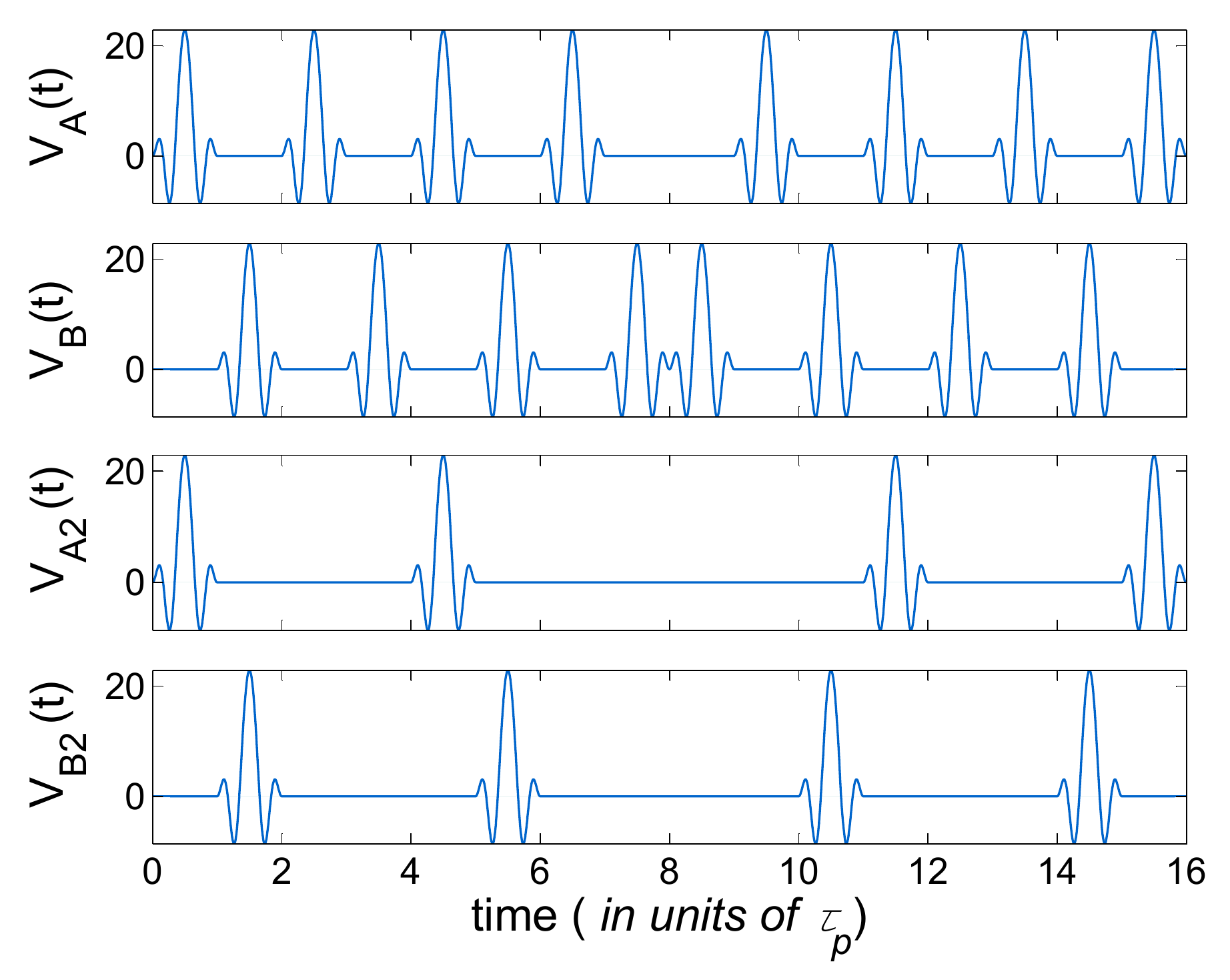}
  \caption{(Color online) Pulse sequences used to implement the
    two-qubit $\exp(-i\alpha\sigma^z_i\sigma^z_j)$ rotations on any
    bipartite graph with equal Ising interactions.  The sequences of
    $\pi$ pulses along the $x-$axis, $V_A(t)$ and $V_B(t)$ are run on
    all idle qubits of the sublattices $A$ and $B$ respectively.
    which decouple the qubit-qubit interactions as well as the
    low-frequency phase noise.  For the qubits to be coupled, we
    replace these with $V_{A2}$ and $V_{B2}$ respectively.  Overall
    this produces an effective Hamiltonian with half the Ising
    coupling remaining only for the chosen pairs of qubits, which
    allows one to implement two-qubit gates.
    The second-order self-refocusing pulses $Q_1(\pi)$ have been
    used in this plot \cite{pryadko-sengupta-2008}.}
  \label{fig:zz}
\end{figure}

The actual soft-pulse implementation of these sequences used in our
simulations is shown in Fig.~\ref{fig:zz}.  We used $\tau_2=0$
and minimum allowed $\tau_1=\tau_p$.  Note that in this particular
implementation the prefactor $f=1/2$ is not adjustable.

\subsection{Other gates}
\label{sec:other-gates}
The constructions described in the previous two sections, the
single-qubit rotations and the adjustable two-qubit $ZZ$ gate, form a
universal set, meaning that an arbitrary unitary transformation in
$n$-qubit Hilbert space can be expressed as their
composition\cite{Barenco-1995}.  In particular, %%% in our simulations, we
%%% needed , which represent $\pi$ rotations
%%% about the $ (\hat x+\hat z)/\sqrt2$ axis;
a single-qubit Hadamard gate can be constructed as a combination of
two rotations:
\begin{equation}
  U^{({\rm H})} = -i
  \exp\Bigl({i\frac{\pi}{4}\sigma^y}\Bigr)\exp\Bigl({i\frac{\pi}{2}\sigma^x}
  \Bigr).
  \label{eq:HADM}
\end{equation}
Each of these can be implemented using a single-qubit DCG construction,
see Sec.~\ref{sec:dcg}.

Similarly, the controlled-not (CNOT) gate can be implemented using the
following identity\cite{Galiautdinov2007,Geller2010},
\begin{eqnarray}
 U_{12}^{(\text{CNOT})}  &=& e^{i\pi/4}\,Y_1X_2\bar X_1\bar Y_1 \bar
 Y_2 \exp\Bigl(-i\frac{\pi}{4}\sigma^z_1\sigma^z_2\Bigr)Y_2\qquad \\
  &=& e^{i\pi/4}\,Z_1 X_2 \bar Y_2
\exp\Bigl(-i\frac{\pi}{4}\sigma^z_1\sigma^z_2\Bigr)Y_2,
\label{eq:CNOT}
\end{eqnarray}
where the gate is applied on the qubit $1$ with the control qubit $2$,
$X_j$ and $Y_j$, $j=1,2$ are the unitaries for single-qubit $\pi/2$
rotations around the corresponding axes, e.g., $X_j \equiv
\exp\Bigl(-i\frac{\pi}{4}\sigma_j^x\Bigr)$, and $\bar X_j$, $\bar Y_j$
are the conjugate rotations.  With the two-qubit $ZZ$ rotation
implemented as $N_\mathrm{rep}$ repetitions of the sequence in
Fig.~\ref{fig:zz}, where the average coupling Hamiltonian is $\bar H
={J\over4}\sigma_1^z\sigma_2^z$, the required time interval is
$\Delta{t}=\pi/J$.  With a single sequence in Fig.~\ref{fig:zz} of
duration $16\tau_p$, this gives the following crucial design equation,
\begin{equation}
J\tau_p = \frac{\pi}{16N_{\rm rep}}.
\label{eq:tau_p}
\end{equation}
Larger values of $N_{\rm rep}$ improve the decoupling accuracy and the
gate fidelity in the limit of low noise, but also increase the cost in
terms of the number of pulses.  For our calculations we used
values of $N_{\rm rep}$ from $1$ to $5$.
%%%  {\color{red}\bf and $\tau_p=1$} We use $\kappa=2$ free
%%% evolution periods, during which time the DD pulses can be judiciously
%%% introduced to decouple neighboring qubits.

Other two-qubit controlled gates such as the controlled-$Z$, C-$Z$,
and controlled-$Y$, C-$Y$, gates can be similarly implemented by
applying suitable transformations to the CNOT (or C-$X$) gate.  We
implemented these using the identities
\begin{eqnarray}
U_{12}^{(\text{C-}Y)} &=& e^{-i\pi/4}\bar X_2\bar Z_1 \bar Z_2
\exp\bigl(-i\frac{\pi}{4}\sigma^z_{1}\sigma^z_{2}\bigr)X_2 ,
\label{CY}\\
U_{12}^{(\text{C-}Z)} &=& e^{-i\pi/4}\bar Z_1 \bar Z_2
\exp\bigl(-i\frac{\pi}{4}\sigma^z_{1}\sigma^z_{2}\bigr) .
\label{CZ}
\end{eqnarray}
Further, two neighboring qubits can be swapped with three CNOT
gates\cite{Barenco-1995}.

We emphasize again that our construction allows parallel execution of
similar gates on sets of qubits which do not share neighboring pairs.
For example, any set of simultaneous single-qubit rotations on the
same sublattice of a bipartite lattice, or simultaneous two-qubit
$ZZ$ rotations between any set of pairs which do not include
neighboring qubits, can be implemented in parallel.

\subsection{Gate characterization}

We verified our analytical  arguments used to build
the quantum gates also by numerical simulations of the single- and the
two-qubit gates.  Specifically, we computed numerically the unitary
evolution matrices $U$ corresponding to each of the pulse sequences
discussed in the previous sections.  The pulses were applied using the
control Hamiltonian~(\ref{eq:control-general}), in the presence of the
Ising couplings~(\ref{eq:ising-network}) and a simplified
time-independent bath~(\ref{eq:ising-bath}) with $B_i=0$ and the
coupling operators $A_i$ replaced by chemical shifts represented by
$c$-numbers, $A_i\to \Delta_i$ [cf.~Eq.~(\ref{eq:chemical-shift})].

\begin{figure}[htbp]
\centering
\includegraphics[width=0.65\columnwidth]{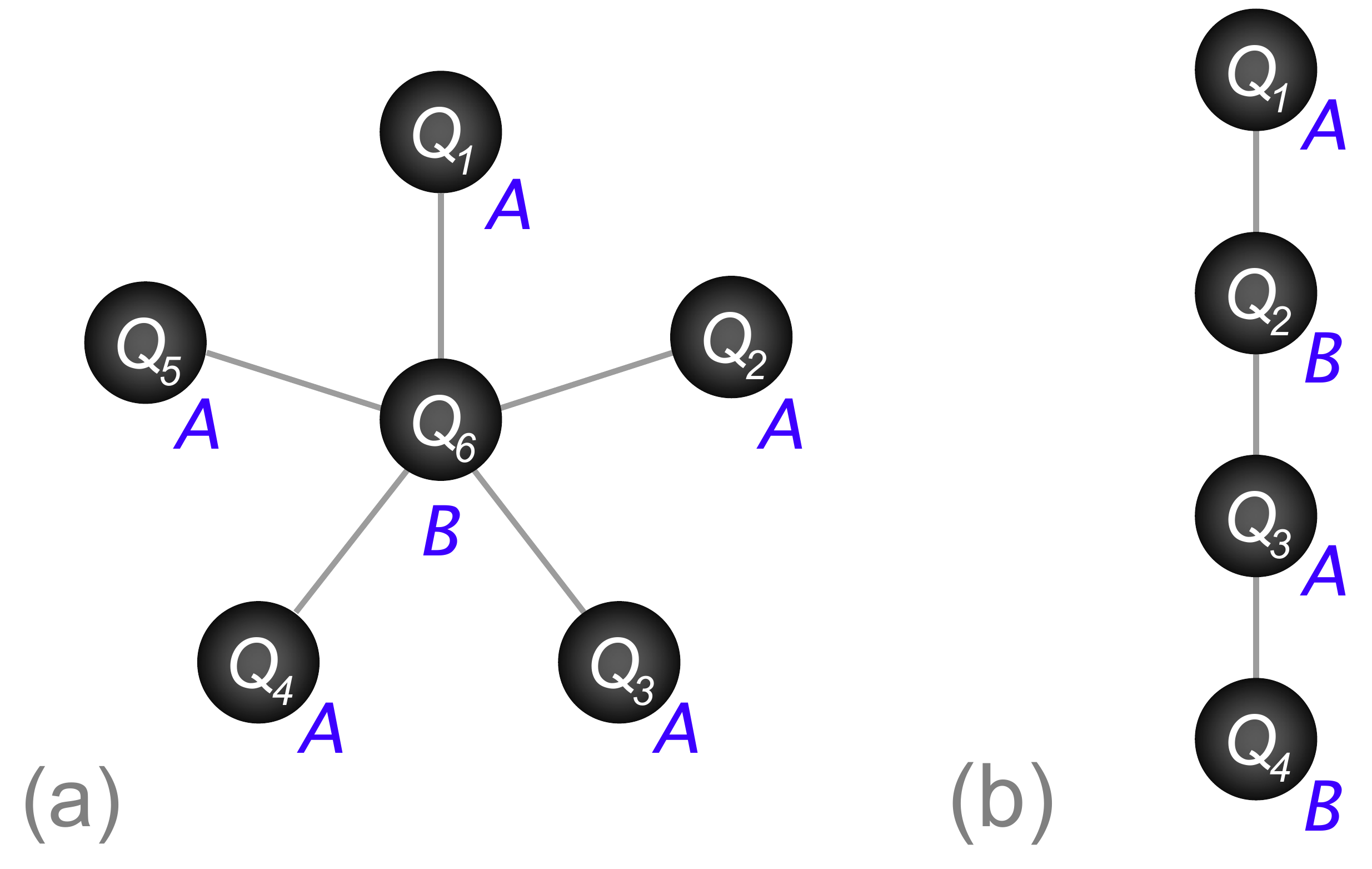}
\caption{(Color online) Two bipartite qubit arrangements (with the
  sublattices as indicated) used for the numerical simulations: (a)
  $n=6$ star graph and (b) $n=4$ chain.  The bonds correspond to Ising
  couplings.}
\label{fig:star}
\end{figure}

Given the ``ideal'' unitary $U_\mathrm{ideal}$ for each gate, we
computed the gate fidelity averaged over initial conditions using the
equation (see the Appendix in Ref.~\cite{pryadko-sengupta-2008})
\begin{equation}
  F(U_\mathrm{ideal},U) =\frac{N+\left|\tr V\right|^2}{N+N^2}, \quad
  V\equiv U_\mathrm{ideal}^{\dagger}U,
\label{eq:Fid}
\end{equation}
where $N$ is the dimension of the Hilbert state, $N=2^n$ for the case
of $n$ qubits.  Specifically, we used two graph families with $n\le6$:
a star graph and a chain, see Fig.~\ref{fig:star}.  In both cases, we
had the Ising coupling Hamiltonians~(\ref{eq:ising-network}) with the
fixed values of the couplings, $J_{ij}=J$.

All simulations have been done with a custom {\tt C++} program using
fourth-order Runge-Kutta algorithm for integrating the unitary
dynamics and the \texttt{Eigen3} library\cite{eigenweb} for matrix
arithmetics.  We used 1024 steps per pulse ($\tau_p$); further
reducing the step size does not improve the accuracy with standard
double precision arithmetics.

\begin{figure*}[htbp]
  \centering
  \includegraphics[width=0.9\textwidth]{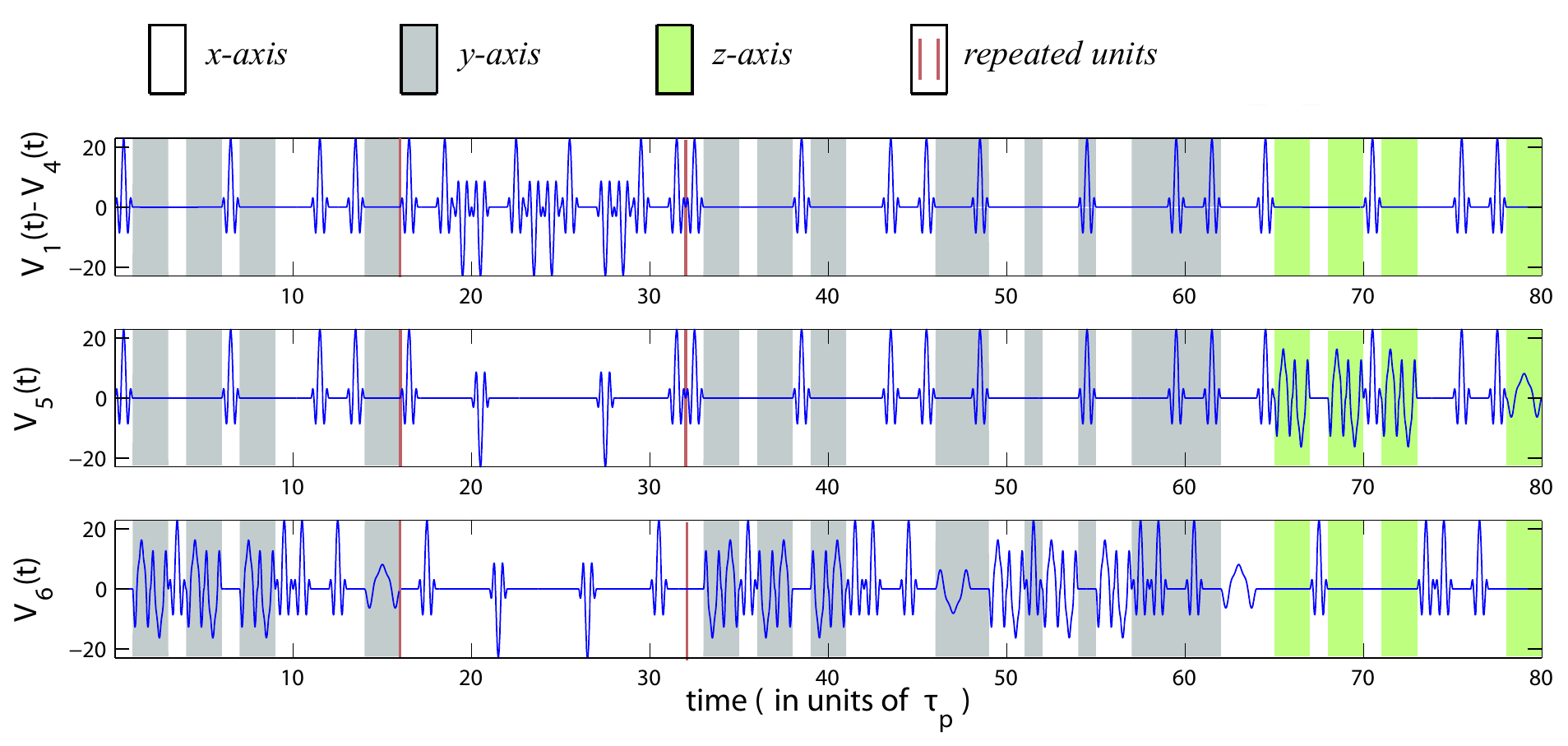}
  \caption{(Color online) Pulse sequences used to implement the CNOT
    gate between qubits $\mathcal{Q}_5$ and $\mathcal{Q}_6$ on a star graph, see
    Fig.~\ref{fig:star}(a).  It is a combination of four DCG gates and
    a $ZZ$-coupling sequence, cf.\ Figs.~\ref{fig:Y3p} and
    \ref{fig:zz}.  Second-order self-refocusing pulse shapes
    $Q_1(\pi)$ and $Q_1(\pi/2)$ from
    Refs.~\onlinecite{sengupta-pryadko-ref-2005,pryadko-sengupta-2008}
    are used.  The shading shows the direction of the applied pulses as
    indicated.  The unit enclosed by vertical red lines,
    $16\tau_p\le t \le 32\tau_p$, should be repeated $N_\mathrm{rep}$
    times, for the total sequence duration
    $16(N_\mathrm{rep}+4)\tau_p$.}
\label{fig:cnot-seqs}
\end{figure*}

Here we discuss the accuracy of the constructed CNOT gate, see
Fig.~\ref{fig:cnot-seqs}.  It is implemented in terms of
$N_\mathrm{rep}$ repetitions of the $ZZ$-decoupling sequence in
Fig.~\ref{fig:zz}, and four single-qubit operations like the one
illustrated in Fig.~\ref{fig:Y3p}, see Eq.~(\ref{eq:CNOT}).  With the
disorder given by chemical shifts only and second-order NMR-style
self-refocusing pulses where $\upsilon=\beta=0$ ($\kappa=\alpha=0$ for
$\pi$-pulses), only the second-order average Hamiltonian $\bar
H^{(2)}$ is non-zero.  This gives the error of the unitary scaling as
$\propto [\max(\Delta_{\rm rms},J)\,\tau_p]^3$, where $\Delta_{\rm
  rms}$ is the r.m.s.\ chemical shift.  The corresponding infidelity
should scale as
\begin{equation}
  1-F\propto [\max(\Delta_{\rm rms},J)\tau_p]^6
  \label{eq:basic-infidelity-scaling}
\end{equation}
on any lattice.  Note that we omitted the dimensionless factors
dependent on the gate duration, $\tau_{\rm
  CNOT}=9*16\,\tau_p=144\,\tau_p$ (the sequences in
Figs.~\ref{fig:Y3p} and \ref{fig:zz} both have duration of
$16\tau_p$), or dependence on the lattice size.  The corresponding
scaling and fault-tolerance of this gate set when used to implement
quantum memory with the toric code are discussed in
Sec.~\ref{sec:scaling}.

\begin{figure}[htbp]
\centering
\includegraphics[width=1\columnwidth]{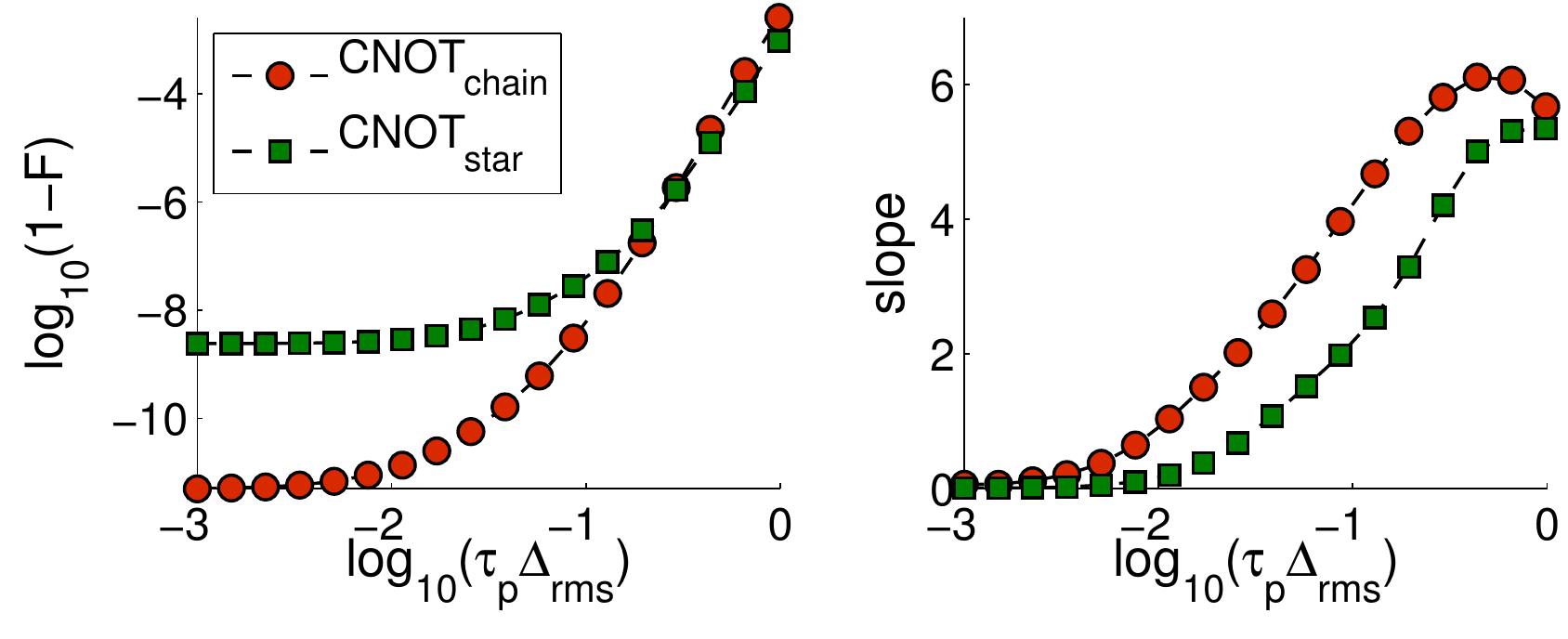}
\caption{(Color online) Comparison of {(a)} average infidelites and
  {(b)} respective slopes for the CNOT gate on an $n=6$ star graph
  vs.\ an $n=6$ chain.  The calculations are averaged over 50 sets of
  random chemical shifts $\Delta_i$ drawn from a zero-average Gaussian
  distribution.}
\label{fig:cnots}
\end{figure}

This scaling~(\ref{eq:basic-infidelity-scaling}) is confirmed in
Fig.~\ref{fig:cnots}, where the infidelities $1-F$ for two lattices as
indicated are plotted on log-log scale as a function of r.m.s.\
chemical shift $\Delta_\mathrm{rms}$.  For larger
$\Delta_\mathrm{rms}$, where the infidelities are dominated by the
chemical shifts $\Delta_i$, the two graphs are very close and they
both have slopes approaching six, in agreement with
Eq.~(\ref{eq:basic-infidelity-scaling}).  Similarly, for small values
of $\Delta_\mathrm{rms}$, the infidelities are dominated by the
decoupling accuracy of the qubit-qubit interactions $J_{ij}$.  Using
variants of the same gate with different $J$ [and different $N_{\rm
    rep}$, see Eq.~(\ref{eq:tau_p})], we verified that in this limit
the infidelity also scales as expected from
Eq.~(\ref{eq:basic-infidelity-scaling}).

\begin{figure}[htbp]
  \centering
  \includegraphics[width=\columnwidth]{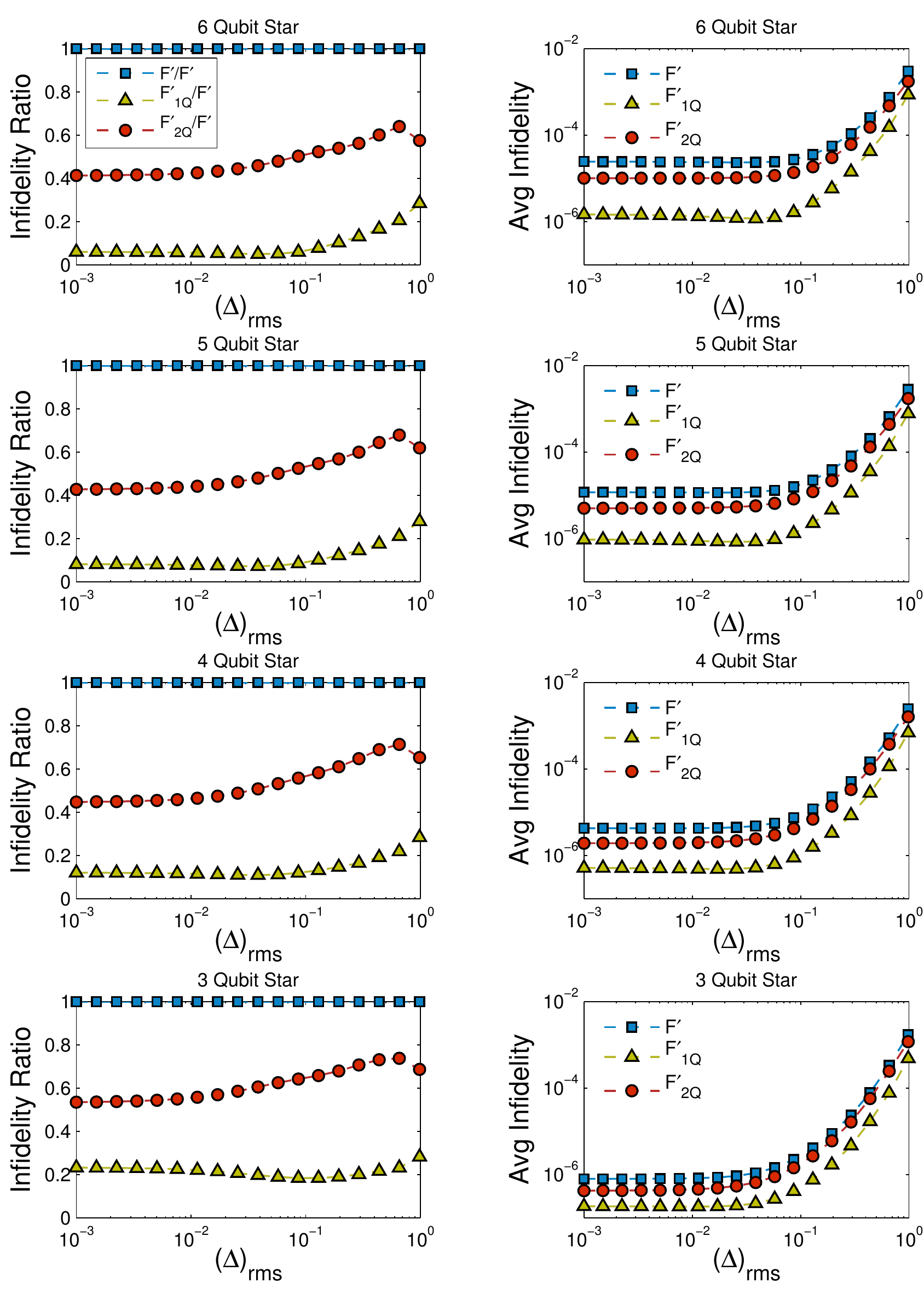}
  \caption{(Color online) Relative (left) and absolute (right)
    contributions of single- and two-qubit errors to the total
    infidelity for CNOT gates implemented on star graphs with
    different numbers of leaves.  Sequences similar to those shown in
    Fig.~\ref{fig:cnot-seqs}
    with $N_\mathrm{rep}=1$ were used to reduce the relative
    contribution of the numerical errors.}
  \label{fig:star-weight}
\end{figure}

Note that a chain where each vertex has at most two neighbors, in the
limit of small $\Delta$ has the infidelity which is smaller by almost
three order of magnitude than that for the star graph of the same
size, $n=6$ [Fig.~\ref{fig:cnots}].  More detailed look into the error
distribution associated with such an increase
in the infidelity is given by Fig.~\ref{fig:star-weight}, where
the relative and absolute contributions of one- and two-qubit errors
to the total gate infidelity are plotted for star graphs with
different numbers of leaves.  To reduce the relative contribution of
the numerical errors, we used sequences similar to those in
Fig.~\ref{fig:cnot-seqs} with $N_\mathrm{rep}=1$.  For small
$\Delta_\mathrm{rms}$, the infidelity is dominated by the errors in
decoupling the inter-qubit couplings.  While for a three-qubit chain
(star $S_2$), one- and two-qubit errors contribute about a quarter
each to the total infidelity in this regime, the contribution of
single-qubit errors drops precipitously with the increased number of
leaves.

Such a dependence is easily explained if we note that the
leading- and subleading-order average Hamiltonians are suppressed in
these calculations, ${\bar H}^{(0)}={\bar H}^{(1)}=0$.  The
contribution of the higher-order terms is dominated by errors of
larger weight: on a star with $z$ leaves, there are ${z\choose 3}$
four-qubit clusters which give contribution to ${\bar H}^{(2)}$,
${z\choose 2}\equiv z(z-1)/2$ three-qubit clusters, and only $z$
two-qubit clusters.  While these terms are strongly suppressed due to
the smallness of $J\tau_p$, in our simulations it is the errors of
weights $w=2$, $3$ and $4$ that are most likely to happen.  In
particular, for $z=5$ ($6$-qubit star) less than 5\% of the total
infidelity for small $\Delta_\mathrm{rms}$ is due to single-qubit
errors.

\begin{figure}[htbp]
\centering
\includegraphics[width=1\columnwidth]{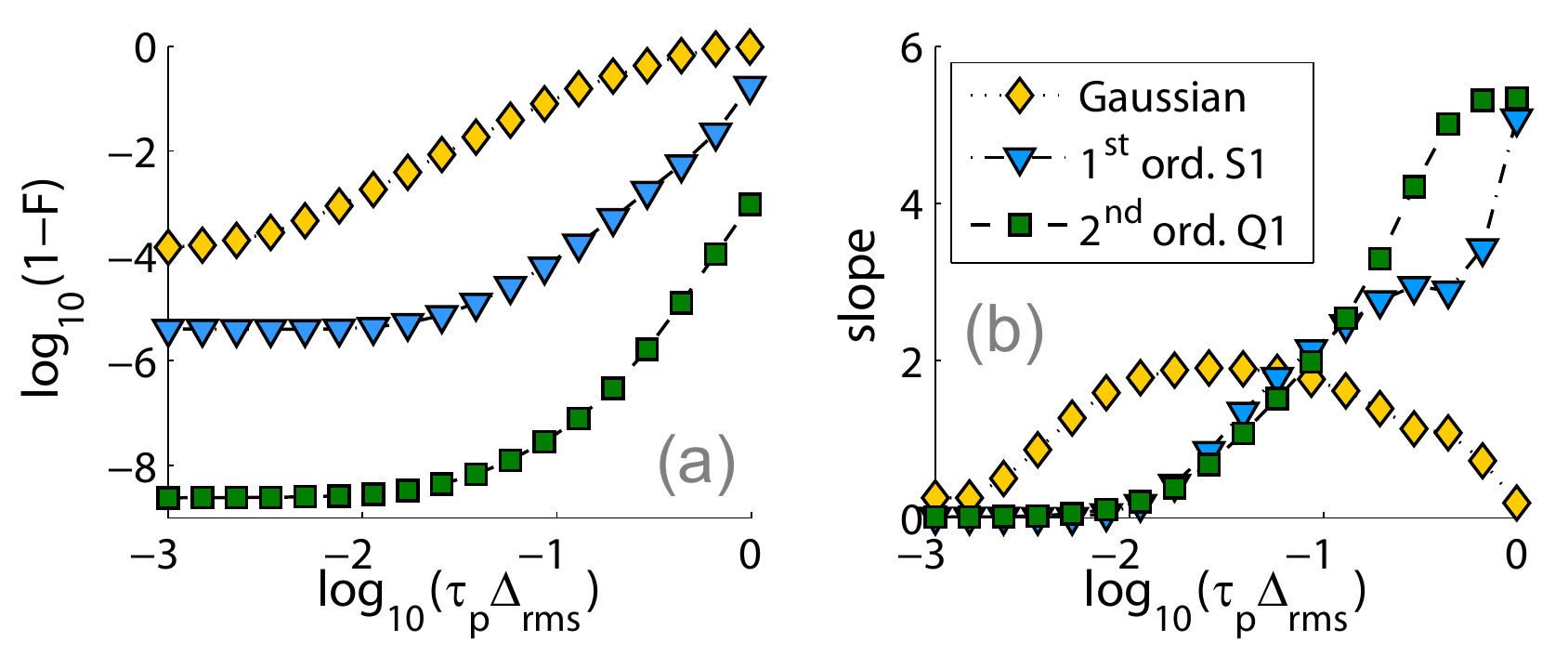}
\caption{(Color online) Comparison of {(a)} average infidelites and
  {(b)} respective slopes for the CNOT gate on an $n=6$ star graph for
  different pulse shapes.  We used Gaussian pulses, $1^{\rm st}$ order
  NMR-type self-refocusing pulses ($S_1$ from
  Ref.~\onlinecite{pryadko-sengupta-2008}) and $2^{\rm nd}$ order
  pulses ($Q_1$ from Ref.~\onlinecite{pryadko-sengupta-2008}).}
\label{fig:dpulses}
\end{figure}

The effect of pulse shape is illustrated in Fig.~\ref{fig:dpulses}.
With first-order pulses, only one coefficient is suppressed,
$\upsilon=0$ ($\kappa=0$ for $\pi$-pulses).  This gives only the
leading-order average Hamiltonian zero ($K=1$\,st order decoupling).
Similarly, with Gaussian pulses, none of the expansion coefficients
introduced in Sec.~\ref{sec:intro-average-ham-pulse} vanishes, so that even
the leading-order effective Hamiltonian is non-zero ($K=0$\,th order
decoupling).  The corresponding unitaries have errors scaling as
$\propto [\max(\Delta,J)\,\tau_p]^{K+1}$ with $K=1$ and $K=0$
respectively, which gives the infidelities $1-F\propto
[\max(\Delta,J)\,\tau_p]^{2K+2}$.  Numerically, we see a dramatic loss
in fidelity associated with these pulses.

\section{Scaling to large system sizes}
\label{sec:scaling}

On star graphs with up to six qubits, and also on chains of different
length (not shown), we saw that for small $\Delta_\mathrm{rms}$ the
infidelity associated with a single CNOT gate is dominated by errors
of weight two and larger, while single-qubit errors are relatively
suppressed.  Such a suppression of few-qubit errors is a typical error
distribution expected with any control scheme relying on decoupling
sequences to remove the unwanted couplings $J_{ij}$.  Indeed, with
finite-duration pulses, generally, one can hope to suppress the
average Hamiltonian only up to some fixed order.  It is the remaining
higher-order terms that are predominantly contributing to multi-qubit
errors.
%%% Moreover, with increasing the total number of
%%% qubits, even if the gate duration remains the same, the weight of
%%% errors likely to happen is going to grow.
An important question is whether such a control scheme can be directly
scaled to large systems.

Superficially, it is difficult to imagine how this can be the case.
Indeed, the coupling Hamiltonian~(\ref{eq:ising-network}) is diagonal,
its spectral norm equals the magnitude of the biggest eigenvalue,
\begin{equation}
  \label{eq:ising-ham-norm}
  \| H_S\| =\sum_{ij} J_{ij}
  \equiv n\langle z J\rangle,
\end{equation}
where $n$ is the total number of qubits, and the second equality
defines the average product of the vertex degree $z$ and the coupling
strength $J$.  Then, even though formally the convergence radius of
the time-dependent perturbation theory is infinite for any finite $n$
and $t$,
with $n$ large, the series is dominated by high
orders which are not easily tractable in this form.

Nevertheless, the pulse-based control scheme can, indeed, be scalable
to large system sizes, when it is combined with an error correcting
code.  Here we only consider the scalability for the specific case of
a toric code implemented on square lattice, with one sublattice used
for ancillae and the other one to encode the state to be protected.
An analysis applicable to more general lattices and quantum
error-correcting codes will be given elsewhere.

\subsection{Decoupling sequence with pulses applied in parallel}
As discussed in the beginning of Sec.~\ref{sec:gate-set}, the analysis
of a collection of pulses simultaneously applied on non-neighboring
sites of an Ising network is simplified by the structure of the
Hamiltonian.  The coupling Hamiltonian in the interaction
representation [see Eq.~(\ref{eq:ising-network})] remains a sum of
commuting terms: bonds connecting the qubits that are not controlled,
and, for every controlled qubit, a ``tuft'' composed of the sum of the
operators for the bonds incident to the corresponding vertex.  The
errors on these two kinds of clusters will involve at most two and
$z+1$ qubits, respectively.  Assuming that the phase errors on sites
and bonds are properly compensated by the sequences, we are left with
the errors due to the individual tuft Hamiltonians with the norm
limited as
\begin{equation}
  \label{eq:tuft-norm-max}
\|  H_\mathrm{tuft}\|=\|  \tilde H_\mathrm{tuft}(t)\|\le {1\over2}zJ.
\end{equation}
The perturbation theory on a single tuft is well controlled when the
expansion parameter
\begin{equation}
  \label{eq:expansion-parameter}
  \alpha_p\equiv {1 \over2}zJ\,\tau_p
\end{equation}
is small.  The norm of $s$-th term in the time-dependent perturbation
theory can be upper-bounded by $\alpha_p^s/s!$, and for $\alpha_p\ll1$
the first non-zero term dominates the expansion.  More precisely, with
order-$K$ self-refocusing pulses, we have the following upper bound on
the total norm of the error operator on a single tuft [cf.\ %
  Eq.~\ref{eq:Fid}],
\begin{equation}
  \label{eq:pulse-error-bound}
d_p\equiv \left\| V-\openone\right\|\le
e^{\alpha_p}-\sum_{s=0}^{K}{\alpha_p^s\over s!}
\le e^{\alpha_p}{\alpha_p^{K+1}\over (m+1)!}.
\end{equation}
When such simultaneous pulse sets are executed repeatedly in a large
system, roughly, $d_p$ is the probability amplitude that an error is
picked up on a given tuft during a given pulse.  At the end of an
error correction cycle, after the measurements of all stabilizer
generators are done, the system is projected to a particular error
configuration with probability given by the square of the sum of all
of the amplitudes which give equivalent errors.  While this
configuration will contain a finite density of errors, for successful
error correction with the toric
code\cite{Dennis-Kitaev-Landahl-Preskill-2002} (as well as generally
for codes with limited-weight stabilizer generators, see
Ref.~\onlinecite{Kovalev-Pryadko-FT-2013} for details) it is important
that these errors do not form large clusters.  Using percolation
theory\footnote{As it turns out, the percolation theory argument for
  existence of a finite threshold in the toric code gives
  qualitatively the same criterion on the distribution of multi-qubit
  correlated errors as that for concatenated codes, see
  Refs.~\protect\onlinecite{Aliferis-Gottesman-Preskill-2006,%
    Aharonov-Kitaev-Preskill-2006}}\label{end:xxx},
the corresponding condition can be written as the
requirement that the total amplitude that a given cluster gets an
error during a single cycle be small,
\begin{equation}
  \label{eq:error-density}
  N_\mathrm{cyc}d_p\ll1.
\end{equation}
With the help of the inequality~(\ref{eq:pulse-error-bound}), this gives
\begin{equation}
  \label{eq:ec-condition-one}
{\alpha_p^{K+1}\over (K+1)!}
  \ll N_\mathrm{cyc}^{-1},
\end{equation}
where we dropped the term $ e^{\alpha_p}$ assuming $\alpha_p\ll1$.

While this is a valid argument, (\textbf{a}) it is only applicable in
the setting of dynamical decoupling, where all terms associated with
the coupling Hamiltonian~(\ref{eq:ising-network}) are suppressed in
the average Hamiltonian, and (\textbf{b}) the upper bound $(z+1)$ on
the typical cluster size contributing to the average Hamiltonian is
too loose and non-specific.  In the following sections, we first
construct a different version of the same argument, looking at
contributions of clusters of different sizes and keeping an accurate
track of their count, and then extend the argument to sequences
forming non-trivial two-qubit gates.

\subsection{Clustering for single-qubit gates}
\label{sec:1qubit-clustering}

Here we consider a typical pulse sequence of duration
$\tau_\mathrm{seq}$ designed to produce a single-qubit gate, or a
collection of single-qubit gates on some subset of qubits.  We assume
a construction similar to the DCGs in Sec.~\ref{sec:dcg}, where the
sequence of a given order $K$ ($K=2$ for the sequence of duration
$\tau_\mathrm{seq}=16\tau_p$ in Fig.~\ref{fig:Y3p}
when second-order NMR-style self-refocusing pulses are used) becomes
exact in the limit of infinitely short pulses, $\tau_p\to0$.
Basically, this means that the full unitary of interest is given
entirely by the non-perturbed unitary $U_0(\tau_\mathrm{seq})$, while
all systematic errors are contained in the slow unitary
$R(\tau_\mathrm{seq})$.

For the toric code, the undetectable errors are formed by products of
same-kind Pauli operators along continuous topologically non-trivial
chains\cite{Dennis-Kitaev-Landahl-Preskill-2002}.  While the error
distribution over (irreducible) clusters is given by the average
Hamiltonian~(\ref{eq:cumulants-defined}), we find it more convenient
to use directly the expansion of the slow evolution operator
$R(\tau_\mathrm{seq})$, see Eq.~(\ref{eq:R-time-ordered}).  Namely, we
further expand each term by writing the interaction Hamiltonian
$\tilde H_i(t_j)$ as a sum of the bond operators.  Generally, each
term in the resulting expansion can be separated into a product of connected
clusters according to which bond operators are present.  The terms in
different clusters always commute and, therefore, the corresponding
time integrations can be rearranged in the
integral~(\ref{eq:R-time-ordered}).  Then, for any decomposition of
the original lattice into a set of disconnected clusters, the
corresponding terms in the expansion of $R(t)$ factor onto a product
of terms corresponding to individual clusters.  The net contribution to a
cluster $Q$ involving $s$ bonds can be written as follows
\begin{equation}
  \label{eq:s-bond-cluster}
  R_{Q}(t)=\sum_{n_i>0} T_t \int_0^t dt_1\ldots\!\!\int_0^t dt_m
  {B_1^{n_1}\over n_1!}{B_2^{n_2}\over n_2!}\ldots {B_s^{n_s} \over n_s!},
\end{equation}
where $B_i^{n_i}$ represents the product of $n_i$ terms for the bond
$i$ evaluated at a subset of time moments $\{t_1,\ldots,t_m\}$, and
$m=n_1+\ldots+ n_s$ is the total number of terms in the product.  The
condition $n_i>0$ is needed to ensure that the entire cluster is covered.
Given the spectral norm for each bond operator, $\|B_i\|=J/2$, we have
the following bound for the contribution of an $s$-bond cluster
\begin{equation}
  \label{eq:s-bond-cluster-norm}
  \|R_{s}(t)\|\le [\exp (t J/2)-1]^s.
\end{equation}
We replaced the subscript $Q$ [see Eq.~(\ref{eq:s-bond-cluster})] with
$s$ since the bound~(\ref{eq:s-bond-cluster-norm}) only depends on the
number of bonds $s$ in the cluster.

Note that for small $tJ$, the obtained expression scales as $\propto
(tJ/2)^s$.  The effect of dynamical decoupling is to suppress any
terms of order $m\le K$ in the expansion of $R(\tau_\mathrm{seq})$.  As
a result, when expanding $R_s(\tau_\mathrm{seq})$, the bound
(\ref{eq:s-bond-cluster-norm}) remains accurate for clusters of size
$s> K$, but there is an additional reduction for small-weight
clusters.  In particular, with $K=2$, the bounds for one- and two-bond
clusters get modified as follows:
\begin{eqnarray}
  \|R_{1}(\tau_\mathrm{seq})\|&\le&
  e^\alpha-1-\alpha-{\alpha^2\over2}\le e^\alpha {\alpha^3\over 6},\\
  \|R_{2}(\tau_\mathrm{seq})\|&\le&
  (e^\alpha-1)^2-\alpha^2\le e^{2\alpha}\alpha^3,
\end{eqnarray}
where $\alpha\equiv \alpha_\mathrm{seq}=\tau_\mathrm{seq} J/2$.  Overall,
for $\alpha\le 1$, we can write the upper bound for the amplitude of a
given $s$-bond cluster as
\begin{equation}
  \label{eq:s-bond-cluster-norm-K}
  \|R_{s}(\tau_\mathrm{seq})\|\le (e\alpha)^{\min(s,K+1)},\quad \alpha\le
  1,
\end{equation}
where $e$ is the base of the natural logarithm; this factor can be
dropped for $\alpha\ll1$.  With this result, an upper bound of the
amplitude that a given point $x$ is in an $s$-bond cluster can be
written as
\begin{equation}
\|M_s\|\le N_s(x)\left\|R_{s}(\tau_\mathrm{seq})\right\|
,\label{eq:amplitude-s-one}
\end{equation}
where $N_s(x)$ is the
number of connected clusters of size $s$ which include the point $x$.

For any regular lattice, the number $N_s$ grows at most exponentially with
$s$, $N_s\le C \mu^s$, where $C>0$ and $\mu>0$ are some constants that
depend on the lattice.  %%% For square lattice we have $\mu\approx
%%% \mathbf{xxx}$;
A general upper bound on $\mu$ for a degree-limited
graph is given by Eq.~(\ref{eq:power-law}).  Overall, for small enough
$\alpha$, this gives an exponential tail of the cluster size
distribution.  Basic conclusion is that errors from parallel single qubit
gates stay local as long as they are executed fast enough.

\subsection{Second interaction Hamiltonian}
\label{sec:2nd-average-ham}

Now, consider a sequence of pulses similar to those in
Figs.~\ref{fig:zz}, \ref{fig:cnot-seqs}, where the leading-order
average Hamiltonian $\bar H_S^{(0)}$ is intentionally non-zero, in
order to implement a part of some multi-qubit gate.  At the same time,
this is an order-$K$ sequence: any correction terms appear only in the
order $K$ and higher of the average Hamiltonian expansion, so that
$\bar H_S^{(m)}=0$, $0<m<K$.

Now, the actual gate has a duration of
$\tau_\mathrm{gate}=N_\mathrm{rep}\tau_\mathrm{seq}$, and we want to
distinguish between the ``wanted'' effect of the leading-order
Hamiltonian $\bar H_S^{(0)}$ and the remaining ``unwanted'' terms
resulting in errors.  To this end, we use the following decomposition
\begin{equation}
  \label{eq:2nd-average-ham-decomposition}
  [R(\tau_\mathrm{seq})]^{N_\mathrm{rep}}=R_0(\tau_\mathrm{gate})\,T_t
  \exp\left(-i\int_0^{\tau_\mathrm{seq}} dt\,\widetilde {\mathop{\delta\!
        H}}(t)\right),
\end{equation}
where $R_0(t)\equiv \exp\biglb(-i t\bar
H^{(0)}(\tau_\mathrm{seq})\bigrb)$ corresponds to the ``wanted''
portion of the unitary generated by the sequence leading-order average
Hamiltonian $H^{(0)}(\tau_\mathrm{seq})$, and $\widetilde
{\mathop{\delta\!  H}}(t)$ is the remaining part of the interaction
Hamiltonian in the interaction representation [see
  Eq.~(\ref{eq:interaction-hamiltonian})], additionally rotated by
$R_0(t)$,
\begin{equation}
  \label{eq:2nd-average-ham-perturbation}
  \widetilde {\mathop{\delta\!  H}}(t)\equiv
  R_0^\dagger(t)\, \left[\tilde H_i(t)-H^{(0)}(\tau_\mathrm{seq} )\right]\, R_0(t).
\end{equation}

We are interested in the specific case where the ``wanted'' unitary is
a product of two-qubit gates on pairs of qubits corresponding to the
edges of the connectivity graph, with each term in the
Hamiltonian~$H^{(0)}(\tau_\mathrm{seq} )$ of the
form~(\ref{eq:avhamAB}), with $|f|\le1$.  Then, the difference
Hamiltonian $\tilde H_i(t)-H^{(0)}(\tau_\mathrm{seq} )$ is a sum of
individual two-qubit bond operators $\tilde B_i$ forming the same
connectivity graph ${\cal G}$, with the norm no more than doubled, $\|
\tilde B_i\|\le J$.  While the unitary
transformation~(\ref{eq:2nd-average-ham-perturbation}) does not change
the norm of individual bond operators, it can change their structure.
A single-qubit operator $\sigma^x$ or $\sigma^y$ on a qubit from a
pair included in $H^{(0)}(\tau_\mathrm{seq} )$ is transformed into a
two-qubit operator; and an Ising bond with one of its qubits driven
can be transformed into a three-qubit operator [we assume that only
  non-neighboring bonds are included in $H^{(0)}(\tau_\mathrm{seq}
  )$].

We can now repeat the arguments from Sec.~\ref{sec:1qubit-clustering}
about the bound~(\ref{eq:amplitude-s-one}) on the total amplitude of
clusters of a given size $s$, connected to a given point $x$.  Namely,
we treat the extended bonds generated by the
transformation~(\ref{eq:2nd-average-ham-perturbation}) as regular
bonds with increased $z$.  On square lattice, this amounts to
increasing from $z=4$ to $z=6$; this nearly doubles the upper bound
for the cluster-number scaling exponent~(\ref{eq:power-law}) to
$\mu_\mathrm{max}\approx 12.21$.  In addition, we have to double the
value of $\alpha_\mathrm{seq}$ to account for possible increased norms
of bond operators; we have $\alpha=J\tau_\mathrm{seq}$.

Now that we have an analog of Eq.~(\ref{eq:amplitude-s-one}) for a
single order-$K$ sequence of duration $\tau_\mathrm{seq}$, we will
estimate errors after $N_\mathrm{rep}$ repetitions of the sequence
simply by scaling the amplitude of each cluster, and using a
percolation-theory argument to account for possible superposition of
different clusters.

An amplitude that a given point is connected to a
size-$s$ cluster is bounded as
\begin{equation}
  \label{eq:any-s-cluster-ampltude}
  \|M_s\|\le N_\mathrm{rep} N_s \|R_s(\tau_\mathrm{seq})\|\propto
  C N_\mathrm{rep} (e \alpha \mu)^{\min (s,K+1)},
\end{equation}
which is exponentially small at large $s$ for $K\ge 1$ and
sufficiently small $\alpha$ since we assume $N_\mathrm{rep}\alpha\sim
N_\mathrm{rep}\tau_\mathrm{seq}J\alt \pi$.   After $N_\mathrm{rep}$
repetitions of the basic sequence, clusters may overlap.  However,
in spite of these overlaps, very large clusters will not form as long
as the cluster density is sufficiently far below the percolation
threshold.

Notice that exponential tail in Eq.~(\ref{eq:any-s-cluster-ampltude})
guarantees the existence of a finite percolation threshold.  Indeed,
an $s$-bond cluster can be always covered with a circle of area $A_s=
\pi\lceil s/2\rceil^2\le \pi s^2$.  For coverage by random circles, a
finite percolation threshold exists iff the radius distribution is
such that the average disk area $\langle A\rangle$ is
finite\cite{Gouere-low-limit-2008}.  Moreover, in a given dimension,
the percolation threshold in terms of the average covered fraction has
a uniform lower bound which depends on the dimension but not on the
details of the radius distribution
function\cite{Gouere-low-limit-2008}.

In our case, we can give the following  upper bound for the average covered
fraction $f\equiv f_\mathrm{gate}$:
\begin{eqnarray}
  \label{eq:amplitude-average-area}
  f&\le& N_\mathrm{rep}\sum_{s=1}^\infty \left\|
      R_s(\tau_\mathrm{seq})\right\| {A_s
      N_s\over s}
  %%% \\ & \le&  C n (\alpha)^{K+1}
  %%% \left[\mu+2\mu^2+\ldots
  %%%     +K\mu^{K}+\mu{\partial\over \partial\mu}{\mu^{K+1}\over
  %%%       1-\alpha\mu} \right].  \nonumber
    \\
    & \le &  \pi C N_\mathrm{rep} (e\alpha)^{K+1}
    \mu{\partial\over \partial\mu}  \left[{\mu(\mu^{K}-1)\over
          \mu-1}+{\mu^{K+1}\over
          1-e\alpha\mu} \right]
    \\
    &\le&   \pi C N_\mathrm{rep}
    (K+1)(e\alpha\mu)^{K+1}\left[{\mu\over (\mu-1)^2}+{1\over
          (1-e\alpha\mu)^2}\right]. \nonumber
\end{eqnarray}
Since $N_\mathrm{rep}\propto\alpha^{-1}$, one needs to ensure at least
first-order decoupling ($K\ge 1$) to be able to scale
$f_\mathrm{gate}$ down under the percolation threshold,
$f_\mathrm{gate}<f_\mathrm{perc}$, and $K\ge 2$ to be able to do it
efficiently.  Once below the percolation threshold, the amplitude to
encounter an error forming a single large cluster becomes
exponentially small.

We note that with small $e\alpha\mu\ll1$, the
series~(\ref{eq:amplitude-average-area}) is dominated by the clusters
of size $s=K+1$; these involve $K+2$ qubits and have r.m.s.\ linear
size of order $s^{1/2}$ which corresponds to area $A_s\sim s$.  With
this estimate,
we can make a somewhat less conservative estimate of the average
covered area fraction~(\ref{eq:amplitude-average-area}),
\begin{equation}
  f_\mathrm{gate}\alt 2C N_\mathrm{rep} (\alpha\mu)^{K+1}, \quad
  \alpha\mu\ll1.
  \label{eq:less-conservative-area-fraction}
\end{equation}

\subsection{Scaling to large system with toric code}

The subsequent discussion requires some familiarity with operation of
the toric code; we recommend
Ref.~\onlinecite{Fowler-Mariantoni-Martinis-Cleland-2012} for an
excellent introduction.

For a toric code implemented on a plane with separate ancillae for
measurement of the plaquette and the vertex stabilizer generators, the
entire measurement cycle can be performed in six basic steps: ancilla
preparation, four CNOT gates, and projective ancilla measurement.
Each ancilla for measuring a product of $Z$ stabilizer generator has
to be prepared in the $\ket0$ state and measured in the $Z$ basis,
while each ancilla for measuring a product of $X$ stabilizer generator
has to be initialized in the $\ket+$ state and measured in the $Z$
basis.

We make rather specific (although not necessarily realistic for every
qubit implementation) simplifying assumptions about the measurement.
Namely, we assume (a) that a projective measurement in the $Z$ basis
can be done near instantaneously, and (b) that after the measurement
the qubit appears in the $\ket0$ or $\ket1$ state according to the
measurement outcome.  The assumption (a) allows us to avoid additional
assumptions about measuring qubits which are coupled, while the
assumption (b) allows to avoid additional assumptions regarding the
ancilla preparation circuit.  Notice that the ancillae need not be
restored to the $\ket0$ state after the measurement.  One limitation
of the present scheme is that CNOT gates can only be executed on pairs
of qubits that do not share neighbors; effectively this doubles the
number of required CNOT gates to eight per measurement cycle.  As a
result, the duration of the entire measurement cycle for the toric
code is the time it takes to execute two Hadamard gates on the
ancillae measuring the $X$-stabilizer generators, and eight CNOT
gates.

With the gates implemented as in Sec.~\ref{sec:gate-set}, a Hadamard
gate has a duration $32\tau_p$, and a CNOT gate
$16(N_\mathrm{rep}+4)\tau_p$.  The overall cycle duration is
\begin{eqnarray}
  \nonumber
  \tau_\mathrm{cyc}&=&2\times 32\tau_p+8\times 16 (N_\mathrm{rep}+4)\tau_p\\
  &=&16 (8 N_\mathrm{rep}+36)\tau_p\le 16\tau_p\times  10N_\mathrm{rep},
  \label{eq:cycle-toric}
\end{eqnarray}
where we assumed $N_\mathrm{rep}\ge 5$.  This implies that the
expected error-covered area fraction computed for a single CNOT gate
[see Eq.~(\ref{eq:amplitude-average-area})] is increased by an
additional order of magnitude.

As a result of the measurement done at the end of each cycle, the
error operator is projected to a state with well-defined stabilizer.
This does not make the error entirely classical as contributions from
the error configurations differing by a product of some stabilizer
generators have to be added coherently (these correspond to all
deformations of error chains with their ends fixed).

Note that while the probabilities of various error configurations are,
as usual, proportional to the magnitude squared of their amplitudes, a
typical outcome will have an error-covered fraction scaling linearly
and not quadratically with the estimate in
Eq.~(\ref{eq:amplitude-average-area}).  Judging from the convergence
of the series, for $\alpha\mu\ll1$, the likely error configuration
will have a spatial structure corresponding to superposition of
randomly placed connected clusters involving up to $K+1$ qubits each,
with the dominant contribution coming from the biggest size.  In the
present model where all of the errors come from incomplete suppression
of the unwanted couplings, see Eq.~(\ref{eq:ising-network}), we expect
to see no correlations between the error patterns encountered in
subsequent measurement cycles.

In the discussed model, the number of the ancilla qubits equals that
of the qubits in the code; the corresponding per-cycle error
probabilities of a qubit error or a measurement error are thus
expected to be equal.  In the absence of correlations, the error
positions can be efficiently recovered from repeatedly measured
syndromes using the minimal matching algorithm, which gives per-cycle
threshold error probability of around $p_c=4\%$ per
qubit\cite{Wang-Fowler-Hollenberg-2011}.  While correlations tend to
favor error chains, with $K=2$, a typical cluster involves four
qubits, and it has the linear size of about two lattice constants.
Simple scaling suggests that the threshold should not be reduced by
more than a factor of four, to about $p_c=1\%$ per qubit per cycle.
Using the area-based estimate $p_c\le 10f_\mathrm{gate}$ [see
  Eq.~(\ref{eq:less-conservative-area-fraction})], with $K=2$, $C=1$, and
$\mu=10$, we obtain the lower bound for the threshold, $\alpha_c\ge
3\times 10^{-4}$, which corresponds to $N_\mathrm{rep}\alt  10^4$.

Note that this bound is loose as we added the amplitudes of all
errors which can happen in the system and have not made any attempt to
account for the reduction in the number of error patterns resulting
from the projective measurement.  While this estimate proves that the
presented universal gate set based on decoupling pulse sequences in a
network of qubits with always-on Ising couplings can in principle be
scalable when used with the toric code, more detailed analysis is
needed to optimize the construction and to establish the actual
threshold.

\section{Conclusions}

In this work we presented the construction and carefully analyzed the
errors associated with the universal gate set based on soft-pulse
dynamical decoupling sequences.  The gates are designed to work on an
idealized network of qubits with always-on Ising couplings forming a
sparse bipartite graph ${\cal G}$.  The construction is based on the
universal gate set presented by us earlier\cite{De-Pryadko-2013}, with the
difference that now they allow for simultaneous two-qubit gates even
in a system where Ising couplings are not identical.

The single-qubit gates are based on the DCG
construction\cite{Khodjasteh-Viola-PRL-2009,Khodjasteh-Viola-PRA-2009};
they allow arbitrary single-qubit rotations.  Any combination of
single-qubit gates can be executed in parallel on non-neighboring
qubits (e.g., the entire sublattice of a bipartite graph).  When
used with second-order NMR-style self-refocusing pulses, the
constructed sequences eliminate the inter-qubit couplings to second
order, and in addition decouple time-independent on-site Ising terms
(chemical shifts) also to second order.  Fluctuating Ising term
(low-frequency phase noise) is decoupled to linear order; second order
decoupling of such terms can also be achieved using a symmetrized
version of the same construction.

The basic two-qubit gate is an arbitrary-angle $ZZ$-rotation; it can
be viewed as a continuous family of doubled Eulerian
sequences\cite{Viola-Knill-2003} which allow flexibility of the
effective coupling: same average rotation rate can be achieved for
qubit pairs with differing Ising couplings.  These gates can also be
executed in parallel on an arbitrary number of qubit pairs with the
restriction that qubits from different pairs cannot be directly
connected to each other.  In addition to providing controlled removal
of unwanted Ising couplings to quadratic order (when used with
second-order NMR-style self-refocusing pulses), these sequences also
decouple low-frequency phase noise to the same order.

We characterized the accuracy of the constructed gates in few-qubit
systems using an extension of the analytical average-Hamiltonian
expansion\cite{pryadko-quiroz-2007,pryadko-sengupta-2008}, and also
numerically by integrating full quantum dynamics of clusters of up to
six qubits in the presence of control pulses, coupling Hamiltonian,
and additional on-site Ising terms.  These simulations confirmed that
the gates are working as designed, with the systematic portion of the
average infidelity of a CNOT gate as small as $10^{-11}$ on a chain
and $10^{-8}$ on an $n=6$ star graph with $N_\mathrm{rep}=5$
repetitions of the basic sequence [see Figs.~\ref{fig:cnot-seqs} and
  \ref{fig:cnots}].

We also went beyond the fidelity and analyzed the
weight distribution of systematic errors generated by our sequences.
It turned out that single- and two-qubit errors are relatively
suppressed, while errors of larger weights dominate the evolution.
Such an error distribution is expected in any control scheme based on
perturbation theory.

Scalable quantum computation being the primary target of the present
construction, we also analyzed the error patterns that would be
expected when this or similarly constructed gate sets are used in a
large system.  It turned out that for sequences suppressing the
inter-qubit couplings to order $K$, when the couplings are small
compared to the inverse sequence duration, dominant errors are formed
by clusters involving up to $K+1$ bonds (up to $K+2$ qubits).  While
such clusters can sometimes merge forming larger-weight errors, we
show that one can choose the parameters so that large error clusters
do not form during a measurement cycle that involves several CNOT and
single-qubit gates.  We analyzed specifically the
measurement cycle of the toric code and the corresponding planar
layout of qubits and ancillae, and demonstrated that fault tolerant
quantum memory can indeed be implemented using our gate set.

A complete analysis of fault-tolerance, e.g., for the toric code, is beyond
the scope of this work.  We notice, however, that the exponential
bound Eq.~(\ref{eq:any-s-cluster-ampltude}) for the amplitude of a large
error clusters is also compatible with the threshold analysis for
concatenated codes with noise that involves long-range temporal and
spatial correlations\cite{Aliferis-Gottesman-Preskill-2006,Aharonov-Kitaev-Preskill-2006}.   Fault-tolerance with a concatenated
code using the present gate set can be demonstrated by choosing a
suitable qubit network, e.g., a linear qubit
chain\cite{Devitt-2004,Fowler-QEC-2004,Fowler-2005}.

The most important parameter that governs the likelihood of a run-away
large-weight error formation is the sparsity of the coupling network.
It can be characterized by the maximum degree $z$ of the corresponding
graph.  On a chain with $z=2$, there are only $s+1$ clusters with $s$
bonds involving a given qubit; with $z>2$, the cluster number grows
exponentially with $s$.  This growth has to be overcome by the small
expansion parameter $\alpha\equiv J\tau_\mathrm{seq}$:  the
amplitude of an error cluster involving $s$ bonds scales  as
$\alpha^s$.

On the other hand, when a large number of qubits are coupled to a
single qubit or other quantum system like a harmonic oscillator, it
would be much more difficult to control the run-away large weight
error formation.  We believe this applies not only to the present gate
set based on decoupling sequences, but generally to any kind of
control scheme where perturbation theory is used, e.g., controlled
coupling schemes based on tuning qubits in and out of resonance.

We wish to thank Kaveh Khodjasteh, Daniel Lidar, and Lorenza Viola for explaining the working of DCGs. We would also like to thank Alexey Kovalev for a number of useful discussions. This work was supported in part by the U.S. Army Research
Office under Grant No. W911NF-11-1-0027, and by the NSF under Grant
No. 1018935.

\appendix
\section{Cluster size distribution}

Here we derive an upper bound on the number of distinct clusters
connected to a given point $x$ on a graph ${\cal G}$ with vertex
degrees limited by $z$.  First, we notice that a size-$s$ cluster
containing $x$ on ${\cal G}$, after cutting any loops, can be mapped
to a size-$s$ cluster on $z$-regular tree ${\cal T}_z$ (Bethe
lattice), with $x$ mapped to the root.  Such a mapping can only
increase the perimeter (size of the boundary, i.e., number of sites
outside the cluster but neighboring with a site inside it).  Any
size-$s$ cluster on $\mathcal{T}_z$ has the perimeter $t_z(s)\equiv
s(z-2)+2$; for a cluster on $\mathcal{G}$ we have $t\le t_z(s)$.

Second,  the number of weight-$s$ clusters which
contain $x$ on ${\cal T}_z$  is\cite{Hu-clustersize-1987}
\begin{eqnarray}
  \nonumber
  N_s&=&  {s z\,[(z - 1) s]!\over s!\, [(z - 2) s + 2]!}  \\
  & =& {s z\over [(z - 2) s + 2][(z - 2) s + 1]}{(z-1)s\choose s}.\quad
  \label{eq:cluster-size-distribution-Tz}
\end{eqnarray}
For large $s$ the binomial can be approximated in terms of the binary
entropy function, $\log_2{n\choose k}=n H_2(k/n)$, $H_2(x)\equiv
-x\log_2(x)-(1-x)\log_2(1-x)$.  The prefactor in
Eq.~(\ref{eq:cluster-size-distribution-Tz}) is smaller than one for
any $s\ge 1$ and  $z>2$; we obtain
\begin{equation}
  \label{eq:power-law}
  N_s\le \mu_\mathrm{max}^s,\quad
  \mu_\mathrm{max}=2^{(z-1)H_2(1/(z-1))},\quad z>2.
\end{equation}
For square lattice %($z=4$) %the exact value is $\mu=...$ while
Eq.~(\ref{eq:power-law}) gives $\mu_\mathrm{max}=27/4= 6.75$.

\bibliographystyle{apsrev}
%\bibliography{bibliography,qc_all,lpp,more_qc,ldpc,more,percol}

\end{document}